\documentclass[12pt,a4paper,fleqn]{scrartcl}
\RequirePackage{amsthm,amsmath,natbib}
\RequirePackage{amssymb,amstext}
\RequirePackage{hypernat}
\usepackage[pdfpagemode=UseOutlines ,plainpages=false
,hypertexnames=false ,pdfpagelabels ,hyperindex=true,colorlinks=true]{hyperref}
	\makeatletter%
	\Hy@breaklinkstrue%
	\makeatother%

\usepackage{subfig}
\usepackage{color}
\definecolor{darkred}{rgb}{0.6,0.0,0.1}
\definecolor{darkgreen}{rgb}{0,0.5,0}
\definecolor{darkblue}{rgb}{0,0,0.5}
\hypersetup{colorlinks ,linkcolor=darkblue ,filecolor=darkgreen
,urlcolor=darkblue ,citecolor=black}

\usepackage{url}
\usepackage{verbatim}
\usepackage{ bbm}
\usepackage{xr}

\renewcommand{\cite}{\citet}
\usepackage{graphicx}

\definecolor{dgreen}{rgb}{0,0.5,0}
\definecolor{dblue}{rgb}{0,0,0.9}
\definecolor{dred}{rgb}{0.6,0.0,0.1}
\definecolor{dgold}{rgb}{0.5,0.3,0.0}
\definecolor{dvio}{rgb}{0.6,0.3,0.5}
\definecolor{gray}{rgb}{0.5,0.5,0.5}

\newcommand{\wh}{\widehat}
\newcommand{\wtl}{\widetilde}
\newcommand{\mb}{\mathbf}

\newcommand{\EE}{{\mathbb E}}
\renewcommand{\P}{\mathbb{P}}
\def\argmin{\mathop{\textrm{argmin}}}
\newcommand{\cT}{{\cal T}}
\newcommand{\cB}{{\cal B}}
\newcommand{\set}[1]{{\left\lbrace #1\right\rbrace }}	
\def\1{\mathop{\mathbbm 1}\nolimits}

\newtheorem{prop}{Proposition}[section]
\newtheorem{coro}[prop]{Corollary}
\newtheorem{theo}[prop]{Theorem}

\newtheorem{lem}[prop]{Lemma}

\newtheorem{example}{Example}[section]

 \newtheorem{assA}{Assumption}

\numberwithin{equation}{section}

   \author{\textsc{Christoph Breunig}\thanks{Department of Economics, Emory University, Rich Memorial Building, Atlanta, GA 30322, USA, e-mail:
\url{	christoph.breunig@emory.edu}}\\
{\small \textit{Emory University}}
  \and\textsc{Enno Mammen}\thanks{Institute for Applied Mathematics, Universit\"at Heidelberg, Im Neuenheimer Feld 205, 69120 Heidelberg, Germany, e-mail: \url{mammen@math.uni-heidelberg.de}}\\
  {\small \textit{Universit\"at Heidelberg}}
  \and  \textsc{Anna Simoni}\thanks{CNRS, CREST - ENSAE - \'{E}cole Polytechnique, 5, Avenue Henry Le Chatelier, 91120 Palaiseau, France, e-mail: \url{anna.simoni@ensae.fr}}\\
  {\small \textit{CREST, CNRS}}\\
}

\title{Ill-posed Estimation in  High-Dimensional Models with Instrumental Variables\thanks{The authors gratefully thank the Co-Editor Oliver Linton, an Associate Editor, and three anonymous referees for their many constructive comments on the previous version of the paper. Financial support by ANR-11-LABEX-0047 and Deutsche Forschungsgemeinschaft through CRC TRR 190 is gratefully acknowledged.}}

\begin{document}
 \maketitle

\begin{abstract}
This paper is concerned with inference about low-dimensional components of a high-dimensional parameter vector $\beta^0$ which is identified through instrumental variables.
We allow for eigenvalues of the expected outer product of included and excluded covariates, denoted by $M$,  to shrink to zero as the sample size increases.
We propose a novel estimator based on desparsification of an instrumental variable Lasso estimator, which is a regularized version of 2SLS with an additional correction term.
This estimator converges to $\beta^0$ at a rate depending on the mapping properties of $M$.
Linear combinations of our estimator of  $\beta^0$ are shown to be asymptotically normally distributed. Based on consistent covariance estimation,
our method allows  for constructing confidence intervals and statistical tests for single or low-dimensional components of $\beta^0$. In Monte-Carlo simulations we analyze the finite sample behavior of our estimator. We apply our method to estimate a logit model of demand for automobiles using real market share data.
\end{abstract}
\begin{tabbing}
 desparsification, ill-posed estimation problem. \\
\end{tabbing}
\begin{tabbing}
\end{tabbing}

\section{Introduction}
In econometric applications, we may want to include a large number of regressors to account for heterogeneity of individuals or simply because economic theory is not explicit about which regressors to include in the model.
These settings often lead to high-dimensional models where the number of parameters to be estimated is close to the sample size or even larger.

In this paper, we consider an instrumental variable (IV) model where the vector of parameters $\beta^0$ is identified through
\begin{align}\label{model:eq}
  Y=X^T\beta^0+ U, \quad \text{ where }\EE[ U Z]=0,
\end{align}
for a scalar dependent variable $Y$, a possibly endogenous vector of covariates $X$, and a vector of instrumental variables and  exogenous covariates $Z$. Our setup is high-dimensional in the sense that the dimension of $\beta^0$ may be larger than the sample size $n$.

This paper is concerned with inference on inner products of $\beta^0$ of the type $a^T\beta^0$ for some vector $a$. In this sense, our model has a semi-parametric interpretation. When a low-dimensional subvector of $\beta^0$ is the parameter of interest and the remaining components of $\beta^0$ are considered as nuisance parameters, then inference on $a^T\beta^0$ implies inference on this low-dimensional subvector of $\beta^0$ for an appropriate choice of the vector $a$. We also allow the subvector of $\beta^0$ of interest to increase slowly with the sample size and provide inference for it. Our main example is when the low-dimensional subvector of $\beta^0$ is associated with endogenous regressors.

As the number of regressors in $X$ may increase with the sample size $n$, also the singular values of the matrix $M$ defined as
\begin{align*}
  M := \EE\left(ZX^T\right),
\end{align*}
depend on $n$. In particular, including additional control variables in the model  might affect the dependence between endogenous regressors and instruments and hence the cross second moment. This leads to situations where the singular values of $M$ decrease with $n$ and the vector $\beta^0$ is thus not strongly identified, following the terminology in \cite{andrews2012estimation}. Also, when the number of endogenous regressors increases with $n$ it is well known that the singular values of $M$ converge to zero in general and might even have an exponential decay. In the high-dimensional case, we then require some form of sparsity of the matrix $M$, i.e., that many entries of $M$ are zero or sufficiently small.

A crucial insight of this paper is to show how the mapping properties of the matrix $M$ affect the asymptotic behavior of our estimator. For instance, we see that the minimal eigenvalue of $M$ slows down the rate of convergence and enlarges the asymptotic variance of our estimator.
In addition, the sparsity pattern of $M$ and the sparsity pattern of the parameter vector $\beta^0$ are shown to be related: less sparsity of $M$ requires a higher degree of sparsity of $\beta^0$ and vice versa. This can be interpreted as an $\ell_1$ analog of the so-called source condition used in the inverse problems literature which links the smoothness of the unknown function to the smoothing properties of the operator that characterizes the inverse problem.

This paper proposes a novel estimation procedure based on a Lasso type estimator, suitably modified to have a tractable limiting distribution for inner products of $\beta^0$. While the Lasso estimator makes use of the underlying sparsity constraints, it is well known that it does not have a tractable limiting distribution. In this paper, we use the methodology of desparsification to make up for this drawback.
Our desparsified IV Lasso estimator for $\beta^0$ corrects the high-dimensional two stage least squares (2SLS) estimator by subtracting a regularization bias. In the case of low dimensions, i.e. under a known sparsity structure, the resulting estimator coincides with the ordinary 2SLS estimator.

We establish the rate of convergence of inner products of our estimator, and show that the rate is affected by the minimum singular value of $M$ (opportunely normalized). In particular, we can show an analog to the nonparametric IV case, where slow rates of convergence are common.
Moreover, inner products of our estimator for $\beta^0$ are shown to be asymptotically normal. The normalization factor for the estimator is shown to be driven by the minimal singular value of $M$. We derive confidence intervals and hypothesis testing procedures for inner products of $\beta^0$. As discussed above, inference results on inner products of $\beta^0$ imply inference results on low-dimensional subvectors of $\beta^0$ or even on subvectors of $\beta^0$ slowly increasing with the sample size. In Monte Carlo simulations, we show that the proposed confidence intervals have accurate size.

It is interesting to note that having the rate of our estimator affected by the minimum singular value of $M$ is similar to what happens for sieve estimation in the nonparametric IV (NPIV) literature. In NPIV literature the rate of convergence is derived under smoothness assumptions of the underlying IV regression functions instead of under sparsity constraints of the IV regression coefficients as in this paper. In particular, model \eqref{model:eq} can be also seen as an approximation of the true relationship between $Y$ and a vector of endogenous covariates based on a dictionary $X$ of transformations of the endogenous covariates. Hence, the two types of assumptions (smoothness and sparsity) provide two alternative frameworks to deal with high-dimension in nonparametric IV regression models.

\paragraph{Related Literature.}
Our paper contributes to the growing literature on inference for structural parameters in sparse high-dimensional IV settings.  Much work in this setting focuses on the case where the dimension of the endogenous variable is small but where there is a large number of available instruments, see \cite{ng2009selecting}, \cite{belloni2012}, and  \cite{BelloniChernozhukovHansen2011}. In this context, \cite{ChaoSwanson2005} and \cite{HansenHausmanNewey2008}, among others, propose methods to account for weak identification when the number of instruments is allowed to be large but not larger than $n$. \cite{HANSEN2014} and \cite{CarrascoDoukali2017} extend this literature by considering cases where the number of weak instruments is larger than $n$ and propose a Ridge regularized jackknife instrumental variable estimation.
When the number of  endogenous regressors in model \eqref{model:eq} is fixed and there are high-dimensional control variables, \cite{ChernozhukovHansenSpindler2015} propose a three step estimator  where high-dimensional  sparse linear models with only exogenous  variables are fitted. In particular, Lasso is only used for the fit of nuisance parameters and the use of the Lasso estimates follows standard lines.
For the fit of the parameters of the endogenous covariates, the criterion function is orthogonalized such that errors in the estimation of the other parameters (i.e. of the nuisance parameters) enter into the model only quadratically. For this reason the classical bounds for the errors of the Lasso estimates of the nuisance parameters suffice. In particular, no debiasing/desparsification of
Lasso estimates is needed at any point of the procedure.  \cite{BelloniChernozhukovFVHansen} consider estimation of treatment effects in IV models with binary instrument and endogenous variable in the presence of a high-dimensional set of control variables.

Also relevant to this paper is the literature concerning choice of valid instruments.  In the context of a scalar endogenous variable, \cite{guo2016} propose a method to select valid instruments based on hard thresholding  in setups where the number of instruments and of
exogenous variables may tend to infinity. Their proposal is related to LASSO approaches
for the selection of valid instruments in finite dimensional setups. \cite{kang2016instrumental}  use Lasso to instrumental variable selection in the context of  invalid instruments. %, i.e., enter the structural equation.
 Based on an initial median estimator, \cite{windmeijer2017} use adaptive Lasso for instrument selection and establish consistency of their procedure.

In model \eqref{model:eq} which allows for increasing dimension of endogenous regressors, \cite{gautier2011high} establish a novel estimation procedure based on novel sensitivity characteristics of the empirical counterpart of $M$ to obtain confidence sets with length depending on the strength of instruments. \cite{gautier2011high} also establish confidence bands after bias correction. \cite{belloni2017simultaneous} use such sensitivity to construct simultaneously valid confidence regions and have proposed a multiplier bootstrap procedure to compute critical values and establish its validity. Their approach is based on orthogonality restrictions when considering linear combinations of the original instruments.\\
\indent Our approach is essentially different from the previous ones as our sparsity condition is based on the population matrix $M$ and not on its empirical counterpart. This allows us to provide a novel link between high-dimensional and NPIV estimation where the first is based on assuming sparsity while the latter is based on assuming smoothness in the underlying model, see \textit{e.g.} \cite{AiChen2003}, \cite{NP03econometrica}, \cite{DFR02}, \cite{Chen08}, and references therein for NPIV estimation. \cite{fan2014} propose a modified Lasso approach for estimation in high-dimensional instrumental variables models. Our paper is also related to \cite{guo2016} and \cite{gold2018} that, as we propose in our paper, use two-step estimators using a threshold procedure and Lasso estimation, respectively, in the first step and desparsification in the second step. However, \cite{gold2018} make assumptions about sparsity that differ from ours and their settings exclude cases where the estimator of components of $\beta$ does not achieve a parametric $\sqrt{n}$-rate. On the other hand, we do allow for singular values of $M$ to tend to zero which yields slower rates and provide novel inference results for inner products of the estimator of $\beta$ of increasing dimension. This is an important feature of our paper as we are thus able to provide an interpretation that is close to the nonparametric IV estimation.
Recently, \cite{neykov2018unified} considered univariate confidence set estimation in a high-dimensional setting but require that the number of instruments coincides with the number of endogenous variables. In contrast, our approach is efficient under sparsity constraints and moreover, it is robust against violations of strong identification, which, as far as we know, has not been addressed in the related literature.

 Our paper is also related to the rich statistical literature on high-dimensional statistical models that contain only exogenous variables and where endogeneity and instrumental variables are not considered, see,  \cite{zhang2014}, \cite{javanmard2014, javanmard2014G} and \cite{vandegeer2014}.
An alternative approach to our desparsified Lasso estimator is ridge regression where an $\ell_2$ penalty is used and the asymptotic distribution results can be readily obtained. This approach in high-dimensional Gaussian regression is considered by \cite{buhlmann2013}. In an extensive simulation study, however, \cite{javanmard2014G} show that the ridge regression approach is overly conservative, which is in line with the theoretical results. This is why we also pursue to desparsify the Lasso estimator rather than using the ridge regression.

The remainder of the paper is organized as follows. In Section \ref{sec:model}, we describe the model, motivate the desparsification procedure and discuss sparsity requirements. Section \ref{sec:inference} contains the rates of convergence and the asymptotic normality results of our estimator. Section \ref{sec:implementation} is concerned with the finite sample performance of our estimator. Simulations and numerical implementation of our inference procedure are in Section \ref{sec:implementation}. In Section \ref{Application} we present an application to demand estimation for automobiles using market share data. All proofs can be found in the appendix.

\paragraph{Notation.}
The $\ell_p$ norm of a vector $a$ is denoted by $\|a\|_p$, $1\leq p\leq \infty$. For a set $S$, the cardinality of $S$ is denoted by $|S|$. For a vector $a$, and $S$ a set of indices, $a_S$ denotes the restriction of $a$ to indices in $S$: $a_S:=\{a_j; j\in S\}$. Further, for a matrix $A$ we use the notation
\begin{align*}
\|A\|_\infty := \max_{j,k} |A_{jk}|
\end{align*}
for the element-wise sup-norm,
\begin{align*}
\|A\|_{op,\infty} := \max_{j} \sum_{k} |A_{jk}|
\end{align*}
for the operator norm, and
\begin{align*}
\|A\|_1 := \max_{k}\sum_j|A_{jk}|
\end{align*}
for the $\ell_1$ norm. For vectors $a$ we have $\|a\|_{op,\infty} = \| a\|_{\infty}$ and for a matrix $A$  it holds
$\|A\|_{op,\infty} = \|A^T\|_{1}.$ The smallest and largest eigenvalue of $A$ are denoted by $\lambda_{\min}(A)$ and $\lambda_{\max}(A)$, respectively. We denote by $A_{j}$ the $j$-th column of the matrix $A$ and by $A_{-j}$ the matrix $A$ without the $j$-th column. We denote by $e_j$ the $j$-th unit column vector. For two positive sequences $a_n$, $b_n$ we use the notation $a_n\sim b_n$ to mean that there are two positive and universal constants $C_1, C_2$ such that $C_1\leq a_n/b_n\leq C_2$. We abbreviate ``with probability approaching one'' to ``wpa1'', and say that a sequence of events $\set{B_n}$ holds wpa1 if $\P(B_n^c) = o(1)$ as $n\rightarrow \infty$.

\section{Model and Methodology}\label{sec:model}
Consider again model \eqref{model:eq}, the high-dimensional instrumental variable model is given by
\begin{align}
  Y=X^T\beta^0+ U\text{ where }\EE[ U Z]=0,\tag{\ref{model:eq}}
\end{align}
where $\beta^0$ is the $ p$--dimensional, unknown parameter of interest. Some of the covariates in $X$ are possibly endogenous in the sense that they are related to the unobservables $U$, i.e.,  $\EE[ UX]$ does not vanish. Here, $Y$ is a scalar dependent variable, $ X$ is a $ p$--dimensional vector of endogenous and exogenous covariates, $Z$ is a $q$--dimensional vector of instrumental variables and  exogenous covariates. So, the vectors $Z$ and $X$ may have elements in common if $X$ contains exogenous covariates.\\
\indent To ensure identification of the parameter $\beta^0$, we assume throughout the paper that $q\geq p$. This condition can be met in at least three situations: (1) the case where one has low dimensional endogenous variables and high-dimensional exogenous controls and the interest is in inference on the coefficients of the low dimensional endogenous variables (examples are: [a] the case where one includes many exogenous controls to account for complex heterogeneity, or [b] the case where one includes many exogenous variables to account for nonlinearities to approximate a partial linear structure); (2) more generally, the case where a high-dimensional linear sieve approach is used to approximate nonlinear and nonparametric instrumental variables models; (3) models that have a rich information in exogenous variation, corresponding to high-dimensional instrumental variables, as for instance in \cite{AngristKrueger1991}.

We also assume throughout the paper that the matrices $M := \EE\left(ZX^T\right)$ and  $\Sigma := \EE\left(ZZ^T\right)$ are of full column rank.
Thus, the parameter vector $\beta^0$ is identified through
\begin{align}\label{beta_0}
\beta^0=(M^T \Sigma^{-1}M)^{-1} M^T\Sigma^{-1}\EE[ZY].
%\beta^0=(M^T \Sigma^{-1}M)^{-1} M^T\Sigma^{-1}\EE[Z^TY].
\end{align}
Estimating $\beta^0$ by simply replacing the matrices on the right hand side by their empirical counterparts fails for two reasons. First, the empirical counterparts of $M$ and $\Sigma$  are in general not of full rank in the high-dimensional case. Second, it is well known that, for large matrices, estimators simply based on the sample mean do not provide satisfactory performance. In this paper, we address these challenges by using regularization procedures.

A common assumption to obtain consistent estimation results in the high-dimensional setting is a sparsity restriction: most of the parameters of $\beta^0$ are zero (exact sparsity) or sufficiently small (approximate sparsity) which implies that
a relatively small number of  regressors in $X$ is sufficient in describing the dependent variable $Y$.

Note that the model is not identified if the minimal eigenvalue of $M^T\Sigma^{-1} M$ is  zero, which we rule out throughout the paper (see Assumption \ref{A:Z}). We thus introduce
\begin{align*}
\omega  := 1/\lambda_{\min}\left(M^T\Sigma^{-1} M\right)
\end{align*}
which satisfies $\omega <\infty$ for each $n\geq 1$ under Assumption \ref{A:Z}. Below, we also assume that the maximal eigenvalue of $M^T\Sigma^{-1} M$ is bounded from above uniformly in $n\geq 1$ and hence,  $\omega $ is strictly positive for all $n\geq 1$.
On the other hand, in many cases we expect that $\omega $ might increase with the sample size $n$ either because the model requires a large number of functions to account for nonlinearity in the endogenous covariates or because the instruments are weak and thus the model is not strongly identified. In the first case, $X^T\beta$ is an approximation of the true nonlinear instrumental regression through approximating functions stored in $X$ whose number increases with $n$. In the second case, weakness of the instruments is captured by close to zero elements in the matrices $M$ and $\Sigma^{-1/2}M$.

Similar to \cite{andrews2012estimation}, we consider the  \textit{strongly identified} case where $\omega $ is uniformly bounded above, and the \textit{semi-strongly identified} case where $\omega $ is unbounded but satisfies $n/\omega \to\infty$. We show below that $\omega $ slows down the rate of convergence of our estimator.
In the semi-strong case, the size of the confidence sets increases relative to $\omega $. There is also a third case which is the \textit{weak identified} case where $n/\omega =O(1)$ but the results of our paper do not apply to it. In this paper, we show that under appropriate assumptions the rate of convergence of our estimator for each component of $\beta^0$ is $\sqrt{\omega /n}$.

Throughout the paper, we assume that a sample $(Y_i,X_i,Z_i)$, $1\leq i\leq n$ of independent and identically distributed copies of $(Y,X,Z)$ is available. We write the vector and matrices of observations as $\mathbf Y = (Y_1,\dots,Y_n)^T$, $\mathbf X=(\mathbf X_1,\dots,\mathbf X_p)$ with $\mathbf X_j=( X_{1,j},\dots,  X_{n,j})^T$ for $1\leq j\leq p$ , and $\mathbf Z=(\mathbf Z_1,\dots,\mathbf Z_q)$ with $\mathbf Z_j=(Z_{1,j},\dots,  Z_{n,j})^T$ for $1\leq j\leq q$. Moreover, the $n$-vector of unobservables is denoted by $\mathbf U = (U_1,\ldots, U_n)^T$. The $(d\times d)$--dimensional identity matrix is denoted by $I_d$ and its $j$--th column by $e_j$.

\subsection{The Desparsified IV Lasso Estimator}\label{sub:sec:est}
In this section, we introduce our estimation procedure which is based on desparsifying a Lasso estimator. The methodology is based on  regularized estimators of the matrices $\Theta:=\Sigma^{-1}$,  $M := \EE[Z X^T]$, and $\Theta^M := (M^T\Theta M)^{-1}$ denoted by $\wh \Theta$, $\wh M$, and $\wh\Theta^M$, respectively, which are defined in Subsection \ref{sub:sec:reg_mat}. We propose the following \textit{desparsified IV Lasso estimator} of $\beta^0$ given by
\begin{align}\label{desp:lasso}
%\wh \beta  =\wtl\beta  + \wh\Theta^M\wh M^T \wh\Theta\,(\mathbf Z^T\mathbf Y/n - \wtl M \wtl\beta),
\wh \beta  =\wh\Theta^M\wh M^T \wh\Theta\mathbf Z^T\mathbf Y/n
  - \big(\wh\Theta^M\wh M^T \wh\Theta\mathbf Z^T\mathbf X/n-I_p\big)\wtl\beta,
\end{align}
where  $\wtl \beta$  is a consistent estimator of $\beta^0$ that makes use of the underlying sparsity assumption. The first summand on the right hand side of \eqref{desp:lasso} corresponds to a regularized empirical analog of $\beta^0$ as in \eqref{beta_0}. The second summand on the right hand side of \eqref{desp:lasso} accounts for the regularization bias of our  matrix estimators.

The proposed estimator naturally extends the 2SLS estimator to the high-dimensional case. Consider the situation of a known sparsity structure where regularization is not required and so $\wh \Theta$, $\wh M$, and $\wh\Theta^M$ are the usual empirical counterparts of $\Theta$, $M$ and $\Theta^M$. In this case it holds $\wh\Theta^M\wh M^T \wh\Theta\mathbf Z^T\mathbf X/n=I_p$ and moreover,  $\wh \beta$ coincides with the 2SLS estimator.

The choice of $\wtl\beta$ is motivated by our sparsity assumption given below and by the asymptotic properties for $\wh\beta$ that we want to obtain. To derive the asymptotic results of our desparsified IV Lasso estimator we make use of the following key decomposition
 \begin{align}\label{eq:dec:key}
  \sqrt{n}(\wh \beta -\beta^0) =  \wh \Theta^M\wh M^T \wh\Theta\,\mathbf Z^T\mathbf U/\sqrt n - \Delta,
\end{align}
for a remainder term $\Delta$ which is given by
\begin{align*}
  \Delta := \sqrt{n}\big(\wh\Theta^M\wh M^T\wh\Theta \mathbf Z^T\mathbf X/n - I_p\big)(\wtl\beta - \beta^0).
\end{align*}
Then, we have to show that $\|\Delta\|_\infty$ is asymptotically negligible under regularity assumptions. In particular, to show this we require that  $\|\wtl\beta - \beta^0\|_1$ is sufficiently small. This property is satisfied by the Lasso estimator and thus we choose $\wtl\beta$ in equation \eqref{desp:lasso} to be the Lasso estimator
which makes use of the underlying sparsity structure imposed on $\beta^0$. Therefore, our estimation procedure is based on the IV Lasso estimator of $\beta^0$ given by
\begin{align}\label{eq:gamma:beta:Lasso}
  \wtl\beta &= \argmin_{\beta\in\mathbb{R}^{p}}\left\{(\mathbf Z^T\mathbf Y/n - \wh M\beta)^T\,\wh\Theta \,(\mathbf Z^T\mathbf Y/n - \wh M \beta) +  2\lambda\,\|\beta\|_1\right\}
\end{align}
for some tuning parameter $\lambda>0$. Then, we replace $\wtl\beta$ in equation \eqref{desp:lasso} to obtain $\wh\beta$.

\subsection{Sparsity Constraints}\label{ss:sparsity:constraints}
In this section we introduce some notations and assumptions about sparsity that we tacitly maintain all along the paper. Let $s_0$ denote the cardinality of the set $S_0$, i.e., $s_0 := |S_0|$, where $S_0$ is a set such that $\|\beta_{S_0^c}^0\|_1$ is sufficiently small. That is, we assume that the set $S_0$ is rich enough such that the parameter vector $\beta^0$ satisfies
\begin{align}\label{approx:sparse:ineq}
\| \beta_{S_0^c}^0\|_1 = \sum_{j \not \in S_0}| \beta _j^0 | \leq  C s_0 \sqrt { \log (p) /n}
\end{align}
for some constant $C>0$. Inequality \eqref{approx:sparse:ineq} imposes approximate sparsity on $\beta^0$: The absolute value of the parameters outside the sparsity set $S_0$ is bounded by some value which tends to zero as the sample size tends to infinity.

Hereafter, we assume that $\Theta$ and $\Theta^M $ exist and assume sparsity with respect to rows of $\Theta:= \Sigma^{-1} $. To this purpose we define
\begin{align*}
s_j := |\set{k\neq j:\, \Theta_{jk}\neq 0}| \qquad \textrm{and} \qquad s_{\max}:=\max_{1\leq j\leq q} s_j.
\end{align*}
The sparsity restriction on $\Theta$ has the following interpretation: if the $(jk)$--th component of $\Theta$ is zero, then the  variables $Z_j$ and $Z_k$ are partially uncorrelated, given the other variables. In particular if $Z$ is jointly normal then we have that  the  variables $Z_j$ and $Z_k$ are
conditionally independent, given the other variables. This also motivates the use of an $\ell_1$--penalty for the estimation of $\Sigma^{-1}$, as we do in Section \ref{sss_Theta} and which was proposed by \cite{meinshausen2006}. Note that it is possible to relax the sparsity constraints but this would lead to a less efficient estimator.

We need to assume some sparsity pattern on $M$, that is, most of the elements in each row or column of $M$ are zero. We conjecture that it would suffice to assume only approximate sparsity for $M$ and $\Theta$ but at the cost of much more technical proofs and notation. For the sparsity of $M$ we introduce the notation
\begin{align*}
s_M := \max_{1\leq k\leq p}|\{j: M_{jk} \neq 0\}|.
\end{align*}
Hence, $\|M\|_1 \leq s_M\|M\|_{\infty}$. For $j=1,\dots,p$, we denote $\gamma_j := \argmin_{\gamma\in\mathbb{R}^{p-1}}\|(\Theta^{1/2}M)_j - (\Theta^{1/2} M)_{-j}\gamma\|_2^2$ and impose the following approximate sparsity condition on $\gamma_j$: we assume there exists a set $S_j$ such that
\begin{align}\label{eq:gamma:s:j}
\| \gamma_{j,S_j^c}\|_1 = \sum_{l \not \in S_j}| \gamma _{jl} | \leq  C  \log  (q) / \sqrt {n}
\end{align}
for some constant $C>0$. Below we also denote $s_j^M := |S_j|$ and for convenience we use the notation $s_{\max}^M := \max_{1\leq j\leq q}s_j^M$.

\subsection{Regularized Matrix Estimators}\label{sub:sec:reg_mat}
In this section, we provide the regularization schemes to construct the approximate inverses $\wh\Theta$ and $\wh\Theta^M$ as well as the regularized estimator $\wh M$. Asymptotic properties of these estimator will be studied in Section \ref{sec:inference}.

\subsubsection{Construction of $\wh\Theta$}\label{sss_Theta} %Since the matrix $\wh\Sigma$ is singular in the high-dimensional case
Here we construct a regularized estimator of the inverse of $\Sigma$ denoted by $\wh\Theta$.
The basic idea to construct such an estimator is to relate the inversion of a $q\times q$ matrix to $q$ regression problems of $\mathbf Z_j$ over $\mathbf Z_{-j}$, where for $1\leq j\leq q$, $\mathbf Z_j=(Z_{1,j},\dots,  Z_{n,j})^T$ is the $j$-th column vector of the matrix $\mathbf{Z}$ and $\mathbf Z_{-j}=(\mathbf Z_{1},\dots,  \mathbf Z_{j-1},\mathbf Z_{j+1},\ldots, \mathbf Z_{q})$.
This approach was introduced by \cite{meinshausen2006} and consists in using the Lasso estimator for nodewise regression. For every $j=1,\ldots, q$ we consider the Lasso estimator
\begin{equation}\label{eq:gamma:U:Lasso}
  \wh\xi_{j} = \argmin_{\xi\in\mathbb{R}^{q-1}}\left\{\|\mathbf Z_j - \mathbf Z_{-j}\xi\|_2^2/n + 2\lambda_j^{\Theta}\|\xi\|_1\right\}
\end{equation}
\noindent for some tuning parameter $\lambda_j^{\Theta}>0$ that will be let to tend to zero as the sample size increases to get asymptotic results. We introduce the $q$-column vector $\wh \Gamma_j=(\wh \Gamma_{kj})_{k=1}^q$ such that
\begin{equation}
  \wh \Gamma_{kj} = \left\{\begin{array}
    {ccl}1& & \textrm{for }k=j\\
    -\wh\xi_{jk}& & \textrm{for }k\neq j
  \end{array}\right.
\end{equation}
with $\wh\xi_{j} = (\wh\xi_{jk})_{k\in\{1,\ldots,q\}\setminus \{j\}}$. By the definition of $\wh \Gamma_j$, it holds $\mathbf Z_j - \mathbf Z_{-j}\wh\xi_j = \mathbf Z \wh \Gamma_j$. Then, the matrix $\wh\Theta = \big(\wh\Theta_{1},\dots, \wh\Theta_{q}\big)^T$ is constructed as
\begin{equation}\label{eq_nodewise_Theta}
  \wh\Theta_{j} = \wh\tau_j^{-2}\wh \Gamma_{j}\quad\text{where}\qquad \wh\tau_j^2 = \|\mathbf Z \wh \Gamma_{j}\|_2^2/n + \lambda_j^{\Theta}\|\wh\xi_{j}\|_1.
\end{equation}
Note that while the population counterpart $\Theta$ is symmetric, its estimator $\wh\Theta$ does not need to be so. For more details on this procedure, we refer to \cite{meinshausen2006}.

\subsubsection{Construction of $\wh M$}\label{sss_M}
A standard sample matrix estimator for the matrix $M$ does not have good performance in the high-dimensional case and regularization is needed. Hence, we propose a thresholding estimator of $M$. Intuitively, we want to eliminate those values of the empirical matrix $\widetilde{M} := \mathbf Z^T\mathbf X/n$ that lie below some specified threshold. More precisely, we propose to use the thresholding estimator $\wh{M} = (\wh{M}_{jk})$ where
\begin{equation}\label{eq_M_est_thresholding}
  \wh{M}_{jk} := \widetilde{M}_{jk} \1\left\{|\widetilde{M}_{jk}| \geq C_0\, \sqrt{\frac{\log q}{n}}\right\}
\end{equation}
\noindent and $C_0 > 0$ is a constant defined as in Proposition \ref{th:norm1:M} and is related to the constant appearing in the large deviation inequality for the components of $\widetilde{M}$. For symmetric matrices such a regularization scheme has been considered in \cite{BickelLevina2008} and \cite{CaiZhou2012} among others, and we refer to these papers for a discussion on this constant. For practical implementation, we choose $c_n := C_0\sqrt{\log(q)/n}$ by cross-validation, see Section \ref{sec:implementation} for more details.

\subsubsection{Construction of $\wh\Theta^M$} In this section, we construct the estimator $\wh\Theta^M$  which is an approximate inverse of $\wh M^T \wh\Theta \wh M$. This estimator involves the regularized estimators $\wh \Theta$ and $\wh M$ obtained in Sections \ref{sss_Theta} and \ref{sss_M} and the square root of $\wh\Theta$, denoted by $\wh \Theta^{1/2}$. Note that we can make use of the Schur decomposition of $\wh\Theta$ to compute its square root given that $\wh\Theta$ is not necessarily symmetric.

Let $(\wh \Theta^{1/2}\wh M)_j$ denote the $j$-th column vector of the matrix $\wh \Theta^{1/2}\wh M$ and
\begin{align*}
(\wh \Theta^{1/2}\wh M)_{-j} := ((\wh \Theta^{1/2}\wh M)_1,\ldots, (\wh \Theta^{1/2}\wh M)_{j-1},(\wh \Theta^{1/2}\wh M)_{j+1},\ldots, (\wh \Theta^{1/2}\wh M)_p).
\end{align*}
Remark that $\wh \Theta^{1/2}\wh M$ is the empirical cross moment of $X$ and the (approximately) orthonormalized $Z$.
The approximate orthonormalization of $\mb Z$ is performed by premultiplication by $\wh\Theta^{1/2}$. As for the construction of $\wh\Theta$, we relate the regularized inversion of a $p\times p$ matrix to $p$ regression problems of $(\wh \Theta^{1/2}\wh M)_j$ on $(\wh \Theta^{1/2}\wh M)_{-j}$. To do that, for every $j=1,\ldots,p$ we consider the Lasso estimator:
\begin{equation}\label{eq:gamma:M:Lasso}
  \wtl\gamma_{j} = \argmin_{\gamma\in\mathbb{R}^{p-1}}\left\{\|(\wh \Theta^{1/2}\wh M)_j - (\wh \Theta^{1/2}\wh M)_{-j}\gamma\|_2^2 + 2\lambda_j^M\|\gamma\|_1\right\},
\end{equation}
for some tuning parameter $\lambda_j^M>0$ that will be let to tend to zero as the sample size increases to get asymptotic results. Let $\wtl \Gamma_j=(\wtl \Gamma_{kj})_{k=1}^p$ be the $p$-column vector determined by
\begin{equation*}
  \wtl \Gamma_{kj} = \left\{\begin{array}
    {ccl}1& & \textrm{for }k=j\\
    -\wtl\gamma_{jk} & & \textrm{for }k\neq j
  \end{array}\right.
\end{equation*}
\noindent with $\wtl\gamma_{j} = (\wtl\gamma_{jk})_{k\in\{1,\ldots,p\}\setminus \{j\}}$.
The matrix $\wh\Theta^M$ is then set equal to $\wh\Theta^M = (\wh\Theta_{1}^M,\dots,\wh\Theta_p^M)^T$ where
\begin{align*}
  \wh\Theta_{j}^M = \wtl\tau_j^{-2}\wtl \Gamma_{j}\qquad \wtl\tau_j^2 = \|\wh \Theta^{1/2}\wh M \wtl \Gamma_{j}\|_2^2 + \lambda_j^M\|\wtl\gamma_{j}\|_1, \qquad 1 \leq j \leq p.
\end{align*}

As already stressed in \cite{vandegeer2014}, other regularization methods to obtain the approximate inverses of $\wh\Sigma$ and $(\wh M^T \wh \Theta \wh M)$ that do not deliver a bound for $\big\|\wh M^T \wh \Theta\wh M\,\wh\Theta_{j}^M-e_j\big\|_\infty$, like the ridge regularization, may not be optimal because without this bound we cannot directly obtain asymptotic distribution results for components of $\beta^0$. The regularization methods that we use to construct $\wh\Theta$ and $\wh\Theta^M$ automatically include this bound in the optimization problem.

\section{Inference}\label{sec:inference}
In this section, we derive the asymptotic distribution of the desparsified IV Lasso estimator $\wh \beta$ given in \eqref{desp:lasso}.
To obtain asymptotic results on which our inference will be based we have to show that the remainder term $\Delta$ in the key decomposition \eqref{eq:dec:key} is asymptotically negligible. We start by providing all the assumptions that we need to obtain our asymptotic results. After that, we first provide results about rates of convergence for the estimated matrices and for $\wh\beta$, and then asymptotic normality will be established.

\subsection{Assumptions}\label{ss:assumptions}
In this section we gather assumptions which we require to establish our inference results. Below, a random vector $W\in\mathbb R^d$ is called sub-Gaussian if $\EE\exp\big(|v^T W|^2/C\big)=O(1)$ for all $v\in\mathbb R^d$ such that $\|v\|_2\leq 1$ and some sufficiently large constant $C>0$.
\begin{assA}\label{A:Z}
  (i) We observe independent and identically distributed (i.i.d.) copies $(Y_1,X_1,Z_1),\dots,(Y_n,X_n,Z_n)$ of $(Y,X,Z)$ satisfying model \eqref{model:eq}.
  (ii)  The vectors $X$ and $Z$ are sub-Gaussian.
  (iii) The eigenvalues of $\Sigma$ are uniformly bounded away from zero and from infinity.
  (iv) The smallest eigenvalue $\lambda_{\min}(M^T\Sigma^{-1}M)$ is bounded from below for each $n\geq 1$ and the largest eigenvalue $\lambda_{\max}(M^T\Sigma^{-1}M)$ is bounded from above uniformly in $n\geq 1$.
\end{assA}
Sub-Gaussianity, as imposed in Assumption \ref{A:Z} $(ii)$, is satisfied, for instance, if the random vectors have bounded support.
Assumption \ref{A:Z} $(iii)$ implies that $\Sigma_{jj} = O(1)$ uniformly in $j$ since $\Sigma_{jj} \leq \lambda_{\max}(\Sigma)$. Similarly, it also implies that $\|\Theta_j\|_2\leq \lambda_{\min}(\Sigma)=O(1)$ uniformly in $j$ and consequently, $\|\Theta\|_1=O\big( \sqrt{s_{\max}}\big)$ which we use below.We also make use of the notation $\cB:=\{\beta:\, \|\beta_{S_0^c}\|_1\leq 3\|\beta_{S_0}\|_1\}$.

\begin{lem}\label{lem:comp}
Let Assumption \ref{A:Z} be satisfied. If $\log(q) /n= o(1)$ then
 it holds  for all $\beta\in\cB$ that
\begin{align}\label{comp:cond:struc}
\|\beta_{ S_0}\|_1^2\leq  s_0\, \beta^TM^T\Sigma^{-1}\widehat\Sigma \Sigma^{-1} M\beta/c^2
\end{align}
wpa1 for some constant $c>0$ and where $\widehat\Sigma = \mathbf Z^T \mathbf Z/n$.
\end{lem}
The previous result shows that a modified version of the so called \textit{compatibility condition}, see \textit{e.g.} \cite{buhlmann2011}, is satisfied with high probability. Note that such conditions are required in the high-dimensional estimation context  in order to relax the requirement of non-zero eigenvalues of associated estimated matrices. The following assumption provides more details about the choice of regularization parameters and imposes conditions on the underlying sparsity.
\begin{assA}\label{A:M}
  (i) It holds $\lambda\sim \log(q)/\sqrt{n}$, $\lambda_j^{\Theta}\sim \sqrt{\log(q)/n}$, and $\lambda_j^M\sim \log(q)/\sqrt{n}$ uniformly in $j$.
  (ii)  It holds $\|M\|_\infty=O(1)$,
  $\EE \big[\max(1,|X^T\beta^0|^2)\|M^T\Sigma^{-1} Z\|_\infty^2\big]=O(\log(p))$
   and $\EE[ U^2|Z]\leq \sigma^2<\infty$ for a constant $\sigma>0$.
 (iii) Assume $s_M\sqrt{s_{\max}}\max( s_{\max}^M,\|\beta^0\|_1)=O\big(\sqrt{\log(q)}\big)$ and
  \begin{align}\label{spars:rate:cond}
s_0s_M\sqrt{s_{\max}}\max\big(\sqrt{s_M}, \sqrt{s_{\max}}\big)\sqrt{\log(p)\log(q)}+ \omega^2s_{\max}^M =o\big(\sqrt{n/\log(q)}\big).
  \end{align}
\end{assA}
Assumption \ref{A:M} $(i)$ specifies the rate of the tuning parameters $\lambda$ used for the plug-in Lasso and $\lambda_j^{\Theta}$, $\lambda_j^M$ used for the nodewise Lasso estimators. The rate of the regularization parameters $\lambda$ and $\lambda_j^M$ is larger by $\sqrt{\log(q)}$ than the common choices of it, which is due to the additional estimation step that is involved for our initial IV Lasso estimator.
 Assumption \ref{A:M} $(ii)$ imposes upper bounds on the maximal element (in absolute value) of $M$ and $M^T\Sigma^{-1} M$, and the conditional variance of $U$ given $Z$, which is standard in the literature and is a mild restriction on the heteroscedasticity of the model.
Assumption \ref{A:M} $(iii)$ imposes sparsity restrictions which we require in order to obtain our inference results. Specifically, this assumption restricts the sparsity of $\beta_0$ (captured by $s_0$) in relation to the sparsity of $M$ (captured by $s_M$). Finally, condition \eqref{spars:rate:cond} implies $\log(p)/\sqrt{n} = o(1)$.

For the next assumption, recall that  $
\gamma_j:= \argmin_{\gamma\in\mathbb{R}^{p-1}}\|(\Theta^{1/2}M)_j - (\Theta^{1/2} M)_{-j}\gamma\|_2^2$ for $1\leq j\leq p$.  Introduce a vector $\Gamma_j:=(\Gamma_{kj})_{k=1}^p$ with $\Gamma_{kj}=-\gamma_{kj}$ for $k\neq j$ and $1$ otherwise, where $\gamma_{kj}$ is the $k$--th entry of $\gamma_j$.
\begin{assA}\label{A:M:tech}
 (i) $\EE \max_{1\leq j\leq p}|(\Theta M\Gamma_j)^TZX^T\Gamma_j|^2=O(\log(p))$ and
$\EE\|M^T\Theta ZX^T\Gamma_j\|_\infty^2=O(\log(p))$ for all $1\leq j\leq p$.
 (ii) It holds
 $\EE \max_{1\leq j\leq p}\|(\Theta M\Gamma_j)^TZ\|_2^4=O(\log(p)^2)$ and further, $\EE \|\Gamma_j^T M^T\Theta Z Z^T\Theta M\|_\infty^2=O(\log(p))$ for all $1\leq j\leq p$.
\end{assA}
Assumption \ref{A:M:tech} $(i)$ imposes upper bounds  on moments associated to $ZX^T$ while  Assumption \ref{A:M:tech} $(ii)$ imposes mild rate conditions on moments of $ZZ^T$. Note that the logarithmic rates in Assumption \ref{A:M:tech} can be replaced
by other powers of logarithms to allow for more heavy tailed variables. This would
require slight changes in our constraints on the growth of dimension parameters $p$ and $q$ and somewhat more restrictive sparsity constraints.

\subsection{Convergence Rates of estimated Matrices}
In this section we provide rates of convergence for the regularized matrices used to construct our estimator $\wh \beta$. These results are then used to establish asymptotic normality results in the next section.

In the following result, we derive a rate of convergence for $\wh M$ in the $\ell_1$ norm. The first part of the theorem provides a large deviation inequality for the components of $\wtl M$ and it is derived by exploiting sub-Gaussianity of the rows of $\mb X$ and $\mb Z$ and a Bernstein-type inequality for sub-exponential random variables.
\begin{prop}\label{th:norm1:M}
  Let Assumption \ref{A:Z} hold. Then, there exists a constant $c>0$ such that
  \begin{equation}\label{lem:result:1}
    \P\big(\big|\widetilde{M}_{jk} - M_{jk}\big| \geq v \big) \leq 4 \exp\big(-c\, v^2n\big)
  \end{equation}
  for $0\leq v < 1$. Moreover, let $\wh M$ be the thresholding estimator defined in \eqref{eq_M_est_thresholding} with $C_0 = \sqrt{8/c}$.
  If in addition Assumption \ref{A:M} \textit{(i)} and \textit{(ii)} holds, then we have
  \begin{align*}
    \big\|\wh M - M\big\|_1 =O_p\big(s_M \sqrt{\log (q)/n}\big).
  \end{align*}
\end{prop}
The constant $c$ in inequality \eqref{lem:result:1} depends on the second order moments and cross moments of the elements in $Z$ and $X$ as well as on their sub-Gaussian norms. Its expression can be deduced from the proof of the proposition given in the appendix.\\
\indent The next result gives a key upper bound for the approximation error of the relaxed inverses $\wh\Theta_j$ and  $\wh\Theta_{j}^M$. These upper bounds depend on the regularization parameters and the values $\wh\tau_j$ or $\widetilde\tau_j$. For the inference on the structural parameter, we thus have to control the asymptotic behavior of  $\wh\tau_j$ and $\widetilde\tau_j$.
\begin{lem}\label{theo:bound}
We have
\begin{align}\label{ineq:kkt}
\big\|\wh\Sigma\wh \Theta_j - e_j \big\|_{\infty} \leq \lambda_j^{\Theta}/\wh\tau_j^2,
\end{align}
and
\begin{align}\label{bound:approx}
\big\|\wh M^T \wh \Theta\wh M\,\wh\Theta_{j}^M-e_j\big\|_\infty \leq \lambda_j^M/\widetilde\tau_j^2.
\end{align}
\end{lem}

We now establish the rate of convergence of the regularized estimators $\wh\Theta$ and $\wh\Theta^M$. The first result in the next proposition was established by \cite{vandegeer2014}, and hence the proof is omitted.
\begin{prop}\label{prop:main}
Suppose Assumption \ref{A:Z} is satisfied.
If $s_0=o(\sqrt{n/\log(q)})$, then we have
\begin{align*}
\|\wh\Theta-\Theta\|_{op,\infty}=O_p\big(s_{\max}\,\sqrt{\log(q)/n}\big).
\end{align*}
If, in addition, Assumptions \ref{A:M} and \ref{A:M:tech} are satisfied then
\begin{align*}
\big\|\wh\Theta^M-\Theta^M\big\|_{op,\infty}=O_p\big(\omega^2s_{\max}^M \log(q)/\sqrt{n}\big).
\end{align*}
\end{prop}

\subsection{Rate of Convergence}
In this subsection, we derive the rate of convergence of the desparsified IV Lasso estimator $\wh \beta$. The next theorem provides an asymptotic upper bound of the bias term $\Delta$, which is key to derive further inference results.
\begin{theo}\label{theo:main}
Let Assumptions \ref{A:Z}--\ref{A:M:tech} be satisfied.
Then, we have
\begin{align*}
\sqrt{n}(\wh \beta-\beta^0)=\omega V+\Delta
\end{align*}
where
\begin{align*}
V=\wh\Theta^M \wh M^T\wh \Theta \mathbf Z^T \mathbf U/(\sqrt n\,\omega)
\end{align*}
and $\Delta$ satisfies
\begin{align*}
\|\Delta\|_\infty=O_p\big(s_0  \, \max(\omega, \|\Theta M{\Theta}^M\|_1) (\log q)^2/\sqrt{n}\big).
\end{align*}
\end{theo}

From Theorem \ref{theo:main} we see that the rate of convergence of the desparsified Lasso estimator $\widehat \beta$ is affected by the possibly increasing parameter $\omega$. In the next result, we show that the bias term $\Delta$ is indeed asymptotically negligible under additional rate requirements. We also see below that the rate of convergence of our estimator is given by $\sqrt{\omega/ n}$ under a mild assumption.
\begin{coro}\label{coro:main}
 Let Assumptions \ref{A:Z}--\ref{A:M:tech} be satisfied. In addition, we assume
\begin{align}\label{coro:cond}
s_0\, (\log q)^2\, \max\big(\sqrt\omega, \|\Theta M{\Theta}^M\|_1/\sqrt\omega\big)= o(\sqrt{ n}).
\end{align}
 Then, we have
 \begin{align*}
\sqrt{n/\omega}(\wh \beta-\beta^0)=\sqrt\omega\,V+o_p(1).
\end{align*}
\end{coro}

We will see in the next section that $V$ after standardization converges to the standard normal distribution and, in particular, that $\sqrt\omega\, V$ is stochastically bounded. We also see that $\omega$ enters the sparsity condition in equation \eqref{coro:cond}.
In the strong identified case, the components of $\wh\beta$ are $\sqrt n$ consistent. In the semi-strongly identified case, this rate of convergence may slow down depending on the asymptotic behavior of $\omega$.

Also the next result is an immediate consequence of Corollary \ref{coro:main} and provides a  bound for linear functionals of  $\wh \beta-\beta^0$ uniformly over representers $a\in\mathbb R^p$ with $\ell_1$ norm which might increase at a rate $K:=K(n)$.
For some constant $C>0$, we define $\mathcal A_K=\set{a\in\mathbb R^p: \,\|a\|_1^2/K\leq C}$.
\begin{coro}\label{coro:uniform}
 Let Assumptions \ref{A:Z}--\ref{A:M:tech} be satisfied. In addition, we assume
\begin{align}\label{coro:uniform:cond}
s_0\, (\log q)^2 \max\big(\sqrt\omega , \|\Theta M{\Theta}^M\|_1/\sqrt \omega\big)= o\big(\sqrt{ n/K}\big).
\end{align}
 Then, we have
\begin{align*}
\sup_{a\in\mathcal A_K}\Big|\sqrt{n/\omega}\, a^T(\wh \beta-\beta^0)-\sqrt\omega\,a^TV\Big|=o_p\big(\sqrt{ K}\big).
\end{align*}
\end{coro}
The sparsity restriction \eqref{coro:uniform:cond} becomes more restrictive for large values of $K$. Two examples of linear functionals for which Corollary \ref{coro:uniform} holds are given by vectors $a$ selecting one component of $\beta$ and vectors $a$ selecting linear combinations of a finite number of components of $\beta$, for which $K=1$ and $K$ is bounded, respectively.
\begin{example}[Series Approximation]
Let $\phi^K(\cdot)$ be a $K$--dimensional vector of basis functions used to approximate a nonlinear relationship between $Y$ and a vector of endogenous variables $X_{\textsl {end}}$ which we assume, in this example, to have bounded support. We assume that  model \eqref{model:eq} holds with $X = \phi^K(X_{\textsl{end}})$.
As basis functions, we consider in this example the Cohen-Daubechies-Vial (CDV) wavelet basis.
Let us denote by $supp(X_{\textsl {end}})$ the (bounded) support of $X_{\textsl {end}}$, then  $\sup_{s\in supp(X_{\textsl{end}})}\|\phi^K(s)\|_1=O(\sqrt K)$ for CDV wavelets, see \cite[Appendix E]{ChenChristensen2015}, which guarantees that $\phi^K(x_{end}) \in \mathcal{A}_K$, for all $x_{end}\in supp(X_{\textsl {end}})$.
If the assumptions of Corollary \ref{coro:uniform} and the rate restriction \eqref{coro:uniform:cond} are satisfied, then Corollary \ref{coro:uniform} yields that
\begin{align*}
\sup_{s\in supp(X_{\textsl{end}}) }\Big|\sqrt{n/\omega} \,\phi^K(s)^T(\wh \beta_{\textsl{end}}-\beta^0_{\textsl{end}})-\sqrt \omega\,\phi^K(s)^T V \Big|=o_p\big(\sqrt{K}\big).
\end{align*}
Consequently, for $\phi^K(s)^T(\wh \beta_{\textsl{end}}-\beta_{\textsl{end}}^0)$ we obtain the rate of convergence $\sqrt{K\omega/ n}$,
 provided that $\sqrt\omega\, \phi^K(s)^TV = O_p(\sqrt K)$ which we establish in the next subsection. This corresponds to the usual variance term in nonparametric IV estimation, see \cite{BCK07econometrica} or \cite{Chen08} and \cite{breunig2016} for pointwise rates. In contrast to the sup-norm convergence results of \cite[Lemma 3.1]{ChenChristensen2015} we do not obtain a $\log(K)$ term since we may exploit sparsity constraints on unknown matrices.
\end{example}

\subsection{Asymptotic Normality}
In this subsection,  we establish asymptotic normality of inner products of the desparsified Lasso estimator $\wh\beta$.
We also see that asymptotic normality of components of $\wh\beta$ immediately follows.

To achieve the asymptotic distribution of our estimator $\wh\beta$ we consider a normalization factor to standardize the estimator $\wh\beta$. This normalization factor involves the empirical counterpart of the covariance matrix of the 2SLS estimator which is given by
\begin{align*}
\Omega = \Theta^M M^T\Theta \EE[U^2 Z Z^T] \Theta M\Theta^M.
\end{align*}
We then require the following assumption on this covariance matrix $\Omega$. We introduce the set $\mathcal A=\set{a\in\mathbb R^p: \,a\in\ell_2\text{ and }\|a\|_1\leq C\|a\|_2}$ for some constant $C>0$.
\begin{assA}\label{A:cov_mat}
There exists a constant $\underline{\sigma}>0$ such that $\sqrt{a^T\Omega a/\omega }\geq \underline{\sigma}\, \|a\|_2$ for all $a\in\mathcal A$.
\end{assA}
Assumption \ref{A:cov_mat} can be easily verified under mild regularity assumptions, such as,  the lower bound $\sqrt{\EE[U^2|Z]}\geq \underline{\sigma}$, which is a common condition to derive asymptotic distribution results. Indeed, the  condition $\sqrt{\EE[U^2|Z]}\geq \underline{\sigma}$ implies $a^T\Omega\,a\geq \underline{\sigma}^2a^T\Theta^M\,a$ and hence, Assumption \ref{A:cov_mat} holds, for instance,  if the eigenvalues of $\Theta^M$ have a polynomial or exponential decay.

We now propose a heteroscedasticity robust covariance estimator. To obtain the empirical counterpart of $\Omega$, denoted by $\wh\Omega$, we replace the matrices $\Theta^M$, $M$, and $\Theta$ by their regularized empirical counterparts defined in Section \ref{sub:sec:reg_mat}:
\begin{align}\label{Omega_hat_def}
\widehat \Omega = n^{-1}\wh\Theta^M \wh M^T\wh \Theta \mathbf Z^T \text{diag}(\wh{ \mathbf U})^2 \mathbf Z\wh\Theta^T \wh M\left(\wh \Theta^M\right)^T,
\end{align}
for the vector of Lasso residuals $\wh{\mathbf U}= \big(Y_1-X_1^T\wtl\beta,\dots, Y_n-X_n^T\wtl\beta\big)$ and $\wtl\beta$ is the IV Lasso estimator given in \eqref{eq:gamma:beta:Lasso}. We now establish  asymptotic normality of linear combinations of the components of $\widehat \beta$.

\begin{theo}\label{th:comp:inf}
Let Assumption \ref{A:cov_mat} and the conditions of Corollary \ref{coro:main} be satisfied. Further, assume that $\max(\EE\|XX^T\|_\infty^2, \EE\|ZZ^T\|_\infty^2)=O(1)$.
Then, for all $a\in\mathcal A$ satisfying
\begin{equation}\label{th:comp:inf:cond}
\omega^{3/2} s_{\max}^M\sqrt{\log(q)}+\sqrt{s_M s_{\max}}\max\big(\sqrt{s_M}, \sqrt{s_{\max}}\big)\|\Theta^M\|_1/\sqrt{\omega} =o\Big(\frac{\sqrt n}{\log(q)}\Big)
\end{equation}
 we have
\begin{align*}
\sqrt{n/(a^T\wh\Omega\, a)}\,a^T\big(\wh \beta-\beta^0\big)\overset{d}{%
\rightarrow} \mathcal{N}(0,1).
\end{align*}
\end{theo}
Below, we provide some implications of Theorem \ref{th:comp:inf}.
An immediate consequence of Theorem \ref{th:comp:inf} is componentwise asymptotic normality, in which case $a=e_j$ for some $1\leq j\leq p$ where $e_j$ is a $p$-vector of zeros but for the $j$-th component that is equal to $1$. Another consequence of Theorem \ref{th:comp:inf} is asymptotic normality of linear combinations of a finite number of components of $\widehat \beta$. In both cases, the restriction imposed in $\mathcal A$ is satisfied. But even if the dimension of the low-dimensional subvector of interest increases, the condition $\|a\|_1/\|a\|_2\leq const.$ can be justified  as the following example illustrates.

\begin{example}[Series Approximation (cont'd)]\label{example:series:approximation}
Let us assume that $\|\phi^K(X_{\textsl{end}})\|_2\sim\sqrt{K}$ almost surely. When a CDV wavelet basis is used, 
recall $\sup_{x\in supp(X_{\textsl{end}})}\|\phi^K(x)\|_1=O(\sqrt K)$. 
For any $x_{\textsl{end}}$ in the support of  $X_{\textsl {end}}$, we may hence assume that the ratio $\|\phi^K(x_{\textsl{end}})\|_1/\|\phi^K(x_{\textsl{end}})\|_2$ is bounded from above uniformly in $n$.
The corresponding sieve variance $\phi^K(x_{\textsl{end}})'\Omega\phi^K(x_{\textsl{end}})$ increases relative to the associated parameter $\omega $ which is thus related to \cite{chen2013} or \cite{ChenChristensen2015}.
\end{example}

The next theorem establishes asymptotically valid confidence intervals and testing procedures for inner products of $\beta^0$. The following two corollaries are direct implications of Theorem \ref{th:comp:inf} and hence, their proofs are omitted. Below, $\Phi$ denotes the cumulative distribution function of the standard normal distribution.
\begin{coro}\label{coro:CI}
  Let the assumptions of Theorem \ref{th:comp:inf} hold. Then, for all $a\in\mathbb{R}^p$ satisfying condition \eqref{th:comp:inf:cond} we have that for any $\alpha\in (0,1)$
\begin{align*}
 \mathbb{P}\left(a^T\beta^0\in\left[a^T\wh\beta \pm \Phi^{-1}(1-\alpha/2)\, (a^T\wh\Omega a)^{1/2}/\sqrt n\right]\right) = 1 - \alpha+o(1).
\end{align*}
\end{coro}
The following examples illustrate the previous theorem for the componentwise case where $a=e_j$.
\begin{example}[Componentwise Confidence Intervals]\label{ex:comp:conf:int}
An asymptotically valid confidence interval for $\beta_{j}^0$ at nominal level $\alpha$ is given by
\begin{align*}
\left[\wh\beta_j-\Phi^{-1}(1-\alpha/2)\, \wh\Omega_{jj}^{1/2}/\sqrt n,\quad \wh\beta_j+\Phi^{-1}(1-\alpha/2)\, \wh\Omega_{jj}^{1/2}/\sqrt n\right].
\end{align*}
 The length of the confidence interval is given by
\begin{align*}
2\Phi^{-1}(1-\alpha/2)\, \wh\Omega_{jj}^{1/2}/\sqrt n.
%=2\Phi^{-1}(1-\alpha/2)\, \Omega_{jj}^{1/2}\frac{\omega }{\sqrt n}
\end{align*}
We thus see that the length of the confidence interval increases relative to the ratio $\sqrt{\omega /n}$. This implies that in the strongly identified case the length of the interval is smaller than in the semi-strongly identified case. If the model is close to be weakly identified then the confidence interval is close to have infinite volume.
This is in line with the findings of \cite{gautier2011high} who showed that in case of weak instruments, confidence sets can be arbitrarily large.
\end{example}

Another direct implication of Theorem \ref{th:comp:inf} concerns hypothesis testing. For some $a\in\mathbb{R}^p$ (satisfying condition \eqref{th:comp:inf:cond}) consider the  null hypothesis $H_{a,0}:\, a^T\beta^0=a^T\beta^H$ for a given vector $\beta^H\in\mathbb{R}^p$.
\begin{coro}
  Let the assumptions of Theorem \ref{th:comp:inf} hold. Then under null hypothesis $H_{a,0}$ we have for any $\alpha\in (0,1)$
\begin{align*}
  \mathbb{P}\left(\frac{\sqrt n \big|a^T(\beta^0-\beta^H)\big|}{\sqrt{a^T\, \wh\Omega\, a}} \geq \Phi^{-1}(1 - \alpha/2)\right) = \alpha+o(1).
\end{align*}
\end{coro}

\section{Numerical Implementation}\label{sec:implementation}
This section presents Monte Carlo experiments to analyze the finite sample properties of our estimator.
We consider the situation where we have a linear reduced form equation but allow for approximate sparsity. We consider three cases: the case where the true structural relationship is linear and we have homoscedasticity (Section \ref{ss_homoscedastic}), the case where the true structural relationship is linear and we have heteroscedasticity (Section \ref{ss_heteroscedastic}), and finally the homoscedastic case where the true structural relationship is nonlinear in the endogenous variable and we use a series approximation (Section \ref{ss_Series_approximation}). All experiments are based on 1000 Monte Carlo iterations.  The choice of tuning parameters is based on $10$-fold cross-validation where we make use of the \texttt{R} function \verb"cv.glmnet" of the \verb"glmnet" package (see Appendix \ref{A:cross:validation} for a description of the cross-validation procedure).

\subsection{Linear structural relationship and homoscedasticity}\label{ss_homoscedastic}
We generate i.i.d. data from the following model
\begin{align}
    Y & = \beta_1 X_1 + \beta_{-1}^T X_{-1}+ U,\qquad X= (X_1, X_{-1}^T)^T, \qquad \beta^0 = (\beta_1,\beta_{-1}^T)^T, \label{simulation:design:1}\\
    X_1 & = \alpha_1 Z_1 + \alpha_{-1}^T X_{-1}+\sqrt{1-\alpha_1^2}\, V,\qquad Z= (Z_1, X_{-1}^T)^T, \qquad \alpha^0 = (\alpha_1,\alpha_{-1}) \nonumber
  \end{align}
with
\begin{align}\label{gen:data}
\left(\begin{array}{c}
  U\\
  V\\
  Z
\end{array}\right) \sim \mathcal{N}\left(0,\left(\begin{array}{ccc}
  1 & \rho & 0\\
  \rho & 1 & 0\\
  0 & 0 & \Sigma\\
\end{array}\right)\right)
\end{align}
where $\Sigma = \left((0.5)^{|j-k|}\right)_{jk}$ is a $q\times q$ matrix. The parameter $\rho$ captures the degree of endogeneity and  is varied in the experiments below.
The parameters are set in the following way: $\beta_1 = 2$, $\beta_{-1,j} = 1 + (j-1)*c$
for $1\leq j \leq 50$, where $c$ is a constant such that the parameters $\beta_{-1,j}$ are equispaced between $1$ and $3$, $\beta_{-1,j} = 0$ for $51 \leq j \leq (p-1)$, and $\alpha_{-1,j} = 	1/(2j^{3})$ for $1\leq j\leq (q-1)$. The parameter $\alpha_1$ accounts for the strength of the instrument $Z_1$ and is varied in the experiments, i.e., we consider $\alpha_1\in\set{1,0.75, 0.5, 0.25}$. Note that we multiply the error term in the second equation by $\sqrt{1-\alpha_1^2}$, to ensure that the variance of $X_1$ does not depend on the value $\alpha_1$. Since $\EE[U^2|Z]=1$ we are in the  homoscedastic case where the covariance matrix simplifies to $\Omega= \EE[U^2]\Theta^M$.\\
\indent The desparsified IV Lasso estimator $\widehat \beta$ is computed as in Subsection \ref{sub:sec:est}. It is based on the initial IV Lasso $\wtl\beta$ given in \eqref{eq:dec:key} where the tuning parameter $\lambda$ is chosen via $10$-fold cross-validation. The regularized estimators $\wh \Theta$, $\wh M$, and $\wh\Theta^M$ are implemented as described in Subsection \ref{sub:sec:reg_mat} with the tuning parameters $\lambda_j^{\Theta}$, $j=1,\ldots,q$, and $\lambda_j^M$, $j=1,\ldots,p$, and $c_n= C_0\sqrt{\log(q)/n}$, chosen by 10-fold cross-validation. We emphasize that our implementation of the estimators for high dimensional matrices follows standard procedures in the related literature see, for instance, \cite{meinshausen2006}. Alternatively, one could use the procedure proposed in \cite{vandegeer2014} and choose the same tuning parameter, say $\lambda_j^{\Theta} = \lambda_{\Theta}$ (resp. $\lambda_j^M = \lambda_M$), by $10$-fold cross-validation among all the $q$ (resp. $p$) nodewise regressions. We examined this procedure but it slows down the computational time and the results were not better. For large choices of $q$ and $p$ we made use of parallel computing (which is straightforward in  \texttt{R} given the \verb"parallel" package).

To estimate the covariance matrix $\Omega$ we adapt to the instrumental variable setting the idea proposed by \cite{sun2012scaled}, which consists in replacing the variance of $U$ by the error variance estimator obtained with the initial IV Lasso $\wtl\beta$, $\wtl\sigma^2 := \sum_{i=1}^n(Y_i - \wtl\beta_1 X_1 - \wtl\beta_{-1}^T X_{-1})^2$. Then, given the estimator $\wh\Omega= \wtl\sigma^2(\hat\Theta^M)^T$ we compute the confidence interval for the structural parameter $\beta_1$ by following  Example \ref{ex:comp:conf:int}.

%\paragraph{Varying $\rho$ and $\alpha$.}
We first study the effect of $\rho$ and $\alpha$ on the results of our inference procedure. Here, we take $p = 100$ with one endogenous variable and $q = 100$ exogenous variables (included and excluded covariates). Then, we look at the effect of $\alpha$ when $p=q=200$. The sample size is fixed to $n=100$. The results are in Table \ref{Table:Bias:Exact_sparsity}. Here, we report the absolute values of the mean bias for the desparsified IV Lasso estimator $\wh\beta_1$ and for the IV Lasso estimator $\wtl \beta_1$, for different values of the parameters $\rho$ and $\alpha_1$. The absolute mean is computed over the $1000$ Monte Carlo replications. We also report the coverage of our confidence interval for $\beta_1$ at the nominal level $95\%$. Table \ref{Table:Bias:Exact_sparsity} also reports the average coverage of the intervals for individual coefficients corresponding to variables in either $S_0$ or $S_0^c$ computed as follows: $AvgCov_{\alpha}(S_0) = s_0^{-1}\sum_{j\in S_0} \wh{\mathbb{P}}(\beta_j^0 \in CI_j(\alpha))$ and $AvgCov_{\alpha}(S_0^c) = (p - s_0)^{-1}\sum_{j\in S_0^c} \wh{\mathbb{P}}(\beta_j^0 \in CI_j(\alpha))$, where $CI_j(\alpha) =[\wh\beta_j \pm \Phi^{-1}(1-\alpha/2)\, \wh\Omega_{jj}^{1/2}/n^{1/2}]$ according to Corollary \ref{coro:CI} and $\wh{\mathbb{P}}$ is obtained as an average over $1000$ Monte Carlo iterations.

\begin{table}[ht!]
\centering
\renewcommand{\arraystretch}{1.3}
  {\small \begin{tabular}{|c|c||c|c|c|c|c|}
    \hline
%Value of
&%Value of
& Absolute & Absolute & Coverage for& Coverage for& Coverage for\\
$\rho$&$\alpha_1$   & mean bias$(\wh\beta_1)$ & mean bias$(\wtl\beta_1)$ & $\beta_1$& $S_0$-coefficients & $S_0^c$-coefficients\\
    \hline
     \hline
\multicolumn{7}{|c|}{$p=q=100$}\\
    \hline
     $0.7$  & $0.75$
& 0.002
& 1.750
& 0.945
& 0.897
& 0.978 \\
  	 & $0.5$ 								
&0.031
& 1.825
& 0.961
& 0.898
& 0.978\\
     & $0.25$ 						& 0.189
& 1.843
& 0.948
& 0.886
& 0.974\\
    \hline
    $0.5$& $0.75$ &0.001
& 1.757
& 0.946
& 0.897
& 0.978 \\
  	 & $0.5$ 							& 0.039
& 1.832
& 0.958
& 0.898
& 0.978 \\
     & $0.25$ 					& 0.220
& 1.863
& 0.944
& 0.884
& 0.974\\
    \hline
       $0.3$ &  $0.75$ & 0.004
& 1.758
& 0.947
& 0.897
& 0.978 \\
  	 & $0.5$ 									& 0.018
& 1.832
& 0.958
& 0.898
& 0.978 \\
     & $0.25$ 							& 0.205
& 1.836
& 0.945
& 0.883
& 0.974\\
     \hline
    \multicolumn{7}{|c|}{$p=q=200$}\\
    \hline
     $0.7$ &  $0.75$ & 0.079
& 1.910
& 0.962
& 0.923
& 0.984\\
  	 & $0.5$ 								& 0.008
& 1.924
& 0.969
& 0.922
& 0.984\\
     & $0.25$ 						& 0.301
& 1.806
& 0.943
& 0.896
& 0.978\\
    \hline
     $0.5$ &   $0.75$ 	& 0.082
& 1.913
& 0.961
& 0.923
& 0.984 \\
  	 & $0.5$ 									& 0.014
& 1.917
& 0.970
& 0.922
& 0.984 \\
     & $0.25$ 							& 0.346
& 1.819
& 0.944
& 0.896
& 0.978\\
    \hline
     $0.3$ &  $0.75$ & 0.087
& 1.913
& 0.962
& 0.923
& 0.984 \\
  	 & $0.5$ 								& 0.026
& 1.917
& 0.969
& 0.922
& 0.984 \\
     & $0.25$ 						& 0.437
& 1.856
& 0.947
& 0.897
& 0.978\\
    \hline
  \end{tabular}}
  \caption{\footnotesize{The simulation design is \eqref{simulation:design:1} with $n = 100$ and varying parameters $\rho$ and $\alpha$. Absolute mean of the bias for the desparsified IV estimator $\wh\beta_1$ and the initial IV Lasso estimator $\wtl \beta_1$. The last three columns provide coverages of our $95\%$-confidence interval for $\beta_1$, and for coefficients corresponding to variables in either $S_0$ or $S_0^c$.}}
\label{Table:Bias:Exact_sparsity}
\end{table}

 From Table \ref{Table:Bias:Exact_sparsity} we see that the absolute mean bias of the desparsified IV Lasso estimator $\widehat \beta_1$ is considerably smaller than the absolute mean bias of the IV Lasso estimator $\widetilde\beta_1$ for each value of $\rho$ and $\alpha_1$, and also as $p=q$ increases.
As $\alpha_1$ decreases, i.e., the strength of instruments declines, we see that the values of the absolute mean bias of both the desparsified IV Lasso and of the initial lasso estimator $\widetilde\beta_1$ become larger when $p=q=100$. When $p=q=200$ we see that the effect on the instrument strength on mean the bias of the IV Lasso estimator $\widetilde\beta_1$ and our desparsified estimator $\widehat \beta_1$ is mixed. 
From the third column of Table \ref{Table:Bias:Exact_sparsity} we see that the empirical coverage for $\beta_1$ is close to the nominal level of $95\%$. %, in particular, when $\alpha_1$ is close to $1$.
 Concerning the coefficients corresponding to variables in $S_0$, we have some undercoverage (see the fourth column of Table \ref{Table:Bias:Exact_sparsity}), which is yet less severe when $p=q=200$. Undercoverage for coefficients in $S_0$ has been shown for the desparsified Lasso in reduced from regression by  \cite{vandegeer2014} in different  simulation designs. On the other hand, the coverage for the coefficients corresponding to variables in $S_0^c$ is accurate and  even somewhat larger than the nominal coverage probability when $p=q=100$.

\begin{figure}[h!]
\captionsetup{format=plain}
   \hspace*{-.02\textwidth}
  \subfloat[\label{fig_Application_1_Overidentified_TP}{}][$\rho = 0.5$, $\alpha=0.25$]{
      \includegraphics[width=0.37\linewidth]{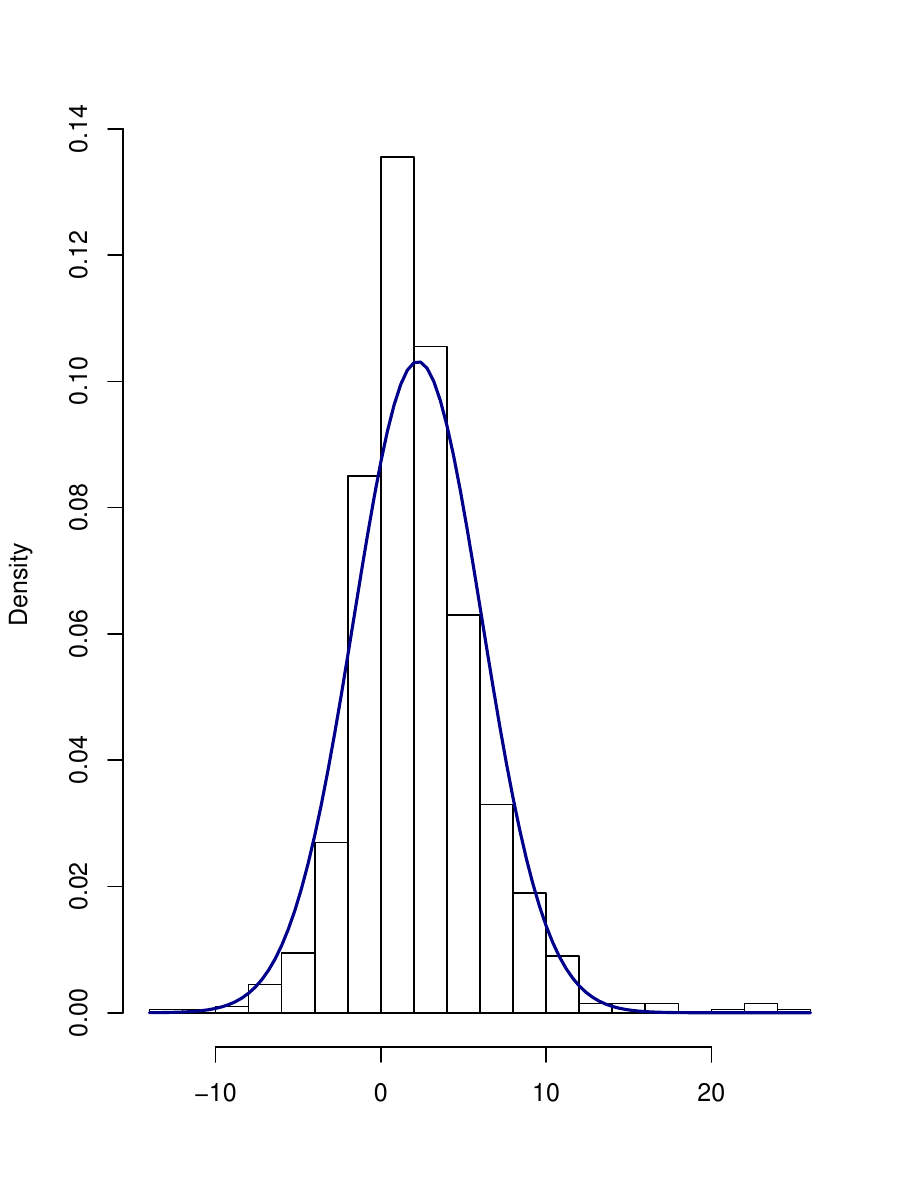}}
  \hspace*{-.045\textwidth}
  \subfloat[\label{fig_Application_1_Overidentified_H}{}][$\rho = 0.5$, $\alpha=0.5$]{
      \includegraphics[width=0.37\linewidth]{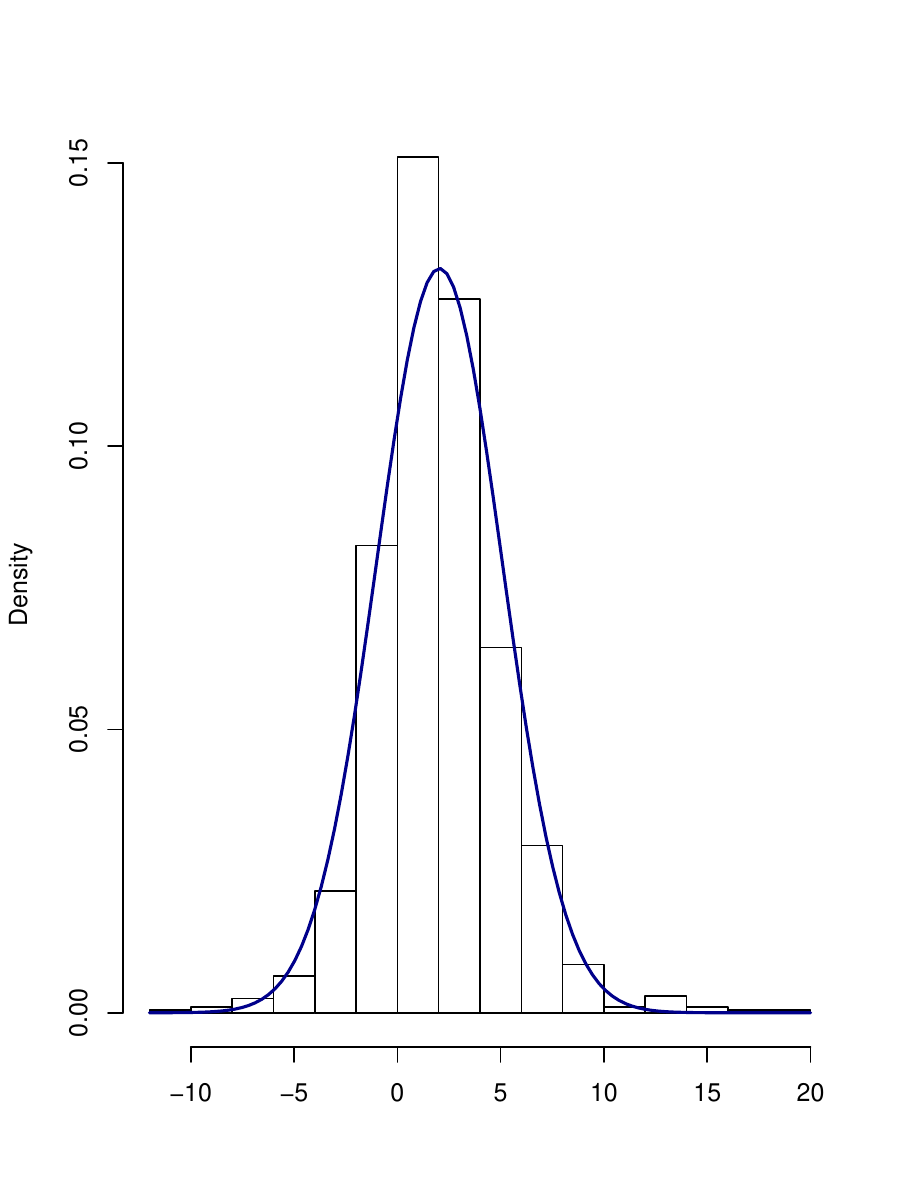}}
 \hspace*{-.045\textwidth}
  \subfloat[\label{fig_Application_1_Overidentified_RM}{}][$\rho = 0.5$, $\alpha=0.75$]{
      \includegraphics[width=0.37\linewidth]{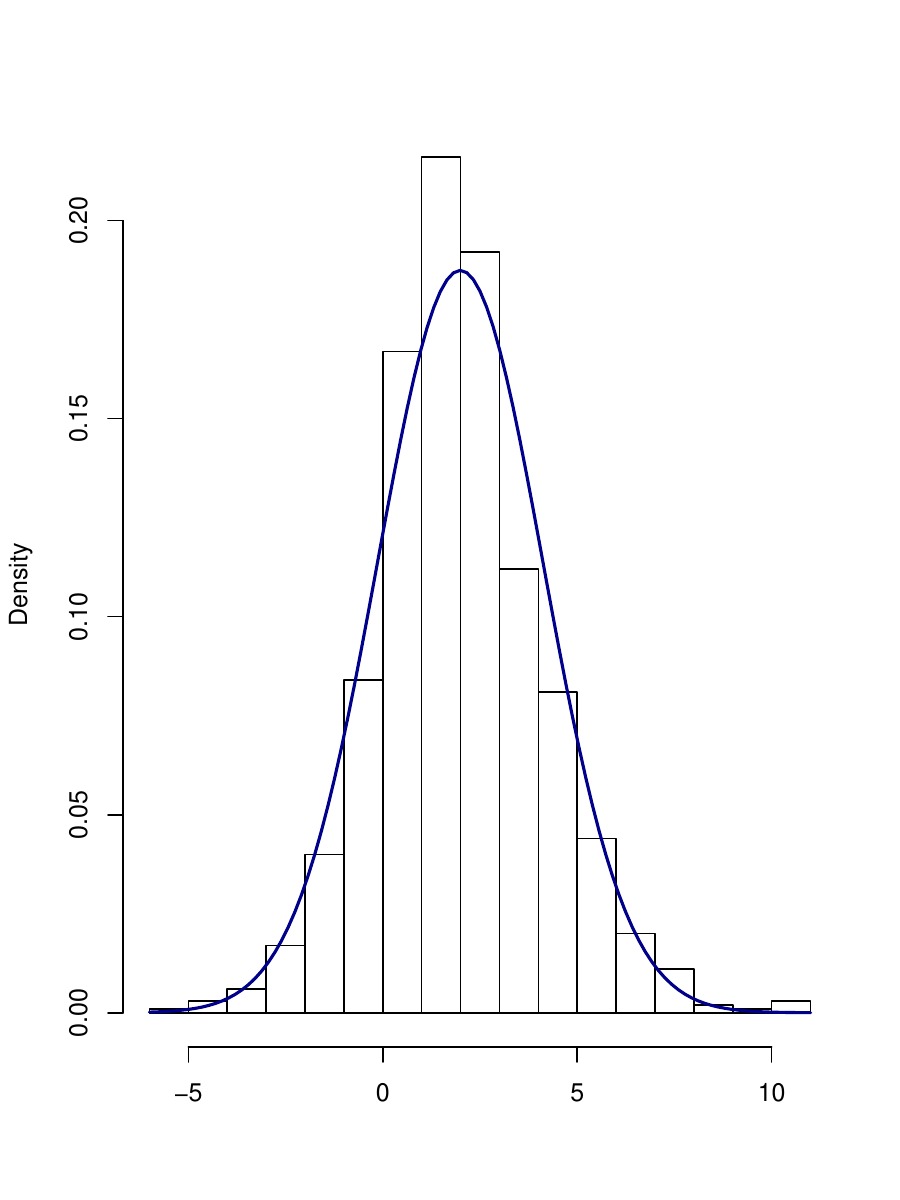}}
  \caption{{\small Histograms approximating the sampling distribution of $\wh\beta_1$ for the simulation design \eqref{simulation:design:1}.}}
  \label{fig_histogram:simulation:1}
\label{Figure:QQ}
\end{figure}

Figure \ref{Figure:QQ} shows the histograms approximating the sampling distribution of our estimator $\wh\beta_1$ for different values of $\alpha_1$ when $\rho=0.5$.
From this figure we see that there is a perfect fit and that for $\alpha_1$ small the distribution is slightly right skewed. We have superposed the probability density function of a standard normal, which corresponds to the asymptotic distribution of the estimator. 

\textbf{Random support of $\beta^0$.} As a further exercise, we have analyzed the situation where the support of $\beta^0$ is randomly selected. We fix the cardinality of the active set of $\beta^0$ equal to $s_0 = 15$ and then the support $S_0$ of $\beta^0$ is obtained as $S_0 = \{u_1,\ldots, u_{15}\}$ where $u_1,\ldots, u_{15}$ is a realization of $15$ draws without replacement from $\{1,\ldots,p\}$. The simulation design is as in  \eqref{simulation:design:1} but where we present here only the result for $\alpha_1=0.05$.  In Table \ref{Table:Bias:Random_support:different:pq} we report the absolute value of the estimated bias (again computed as the difference between the average over the Monte Carlo iterations and the true value of $\beta_1$) for our desparsified IV Lasso estimator $\wh\beta_1$ and for the IV Lasso estimator $\wtl\beta_1$. We see that the coverage of our confidence interval for $\beta_1$ is close to the nominal coverage of $95\%$.
Table \ref{Table:Bias:Random_support:different:pq} also reports the average coverage of the intervals for individual coefficients corresponding to variables in either $S_0$ or $S_0^c$. Again there is some undercoverage for the $S_0$ coefficients.

\begin{table}[ht!]
\captionsetup{format=plain}
\centering
\renewcommand{\arraystretch}{1.3}
 {\small \begin{tabular}{|c||c|c|c|c|c|}
    \hline
    & Absolute & Absolute & Coverage for & Coverage for & Coverage for\\
    & mean bias$(\wh\beta_1)$ & mean bias$(\wtl\beta_1)$ & $\beta_1$ & $S_0$-coefficients & $S_0^c$-coefficients\\
    \hline\hline
 $p=q=100$ & 0.489
&1.802
& 0.941
&0.7741
&0.9753\\
    \hline
 $p=q=150$ &0.401
&1.869
&0.942
&0.691
&0.979\\
 \hline
 $p=q=200$ & 0.505
&1.928
&0.936
& 0.603
&0.981\\
 \hline
  \end{tabular}}
  \caption{\footnotesize{Random support for $\beta^0$ for varying $p$ and $q$. Absolute mean of the bias for the desparsified IV estimator $\wh\beta_1$ and the initial IV Lasso estimator $\wtl \beta_1$. The last three columns provide coverages of our $95\%$-confidence interval for $\beta_1$, and for coefficients corresponding to variables in either $S_0$ or $S_0^c$ when $n=100$.}} \label{Table:Bias:Random_support:different:pq}
\end{table}

%============================================================================================================================
\subsection{Linear structural relationship and heteroscedasticity}\label{ss_heteroscedastic}

We generate i.i.d. data from the model \eqref{simulation:design:1} where
\begin{align}\label{simulation:design:heteroscedasticity}
U = \varepsilon \sqrt{1/2 + \Phi(X_1)}
\quad \text{ and }\quad
\left(\begin{array}{c}
  \varepsilon\\
  V\\
  Z
\end{array}\right) \sim \mathcal{N}\left(0,\left(\begin{array}{ccc}
  1 & 0.5 & 0\\
  0.5 & 1 & 0\\
  0 & 0 & \Sigma\\
\end{array}\right)\right)
\end{align}
where $\Sigma$ is a $q\times q$ matrix that can be set in two different ways. Denote $p_z = q - p_w$, $q = \dim(Z)$ and $p_w = \dim(X_{-1})$, then we have the two following designs for $\Sigma$. 
\begin{itemize}
\item[] \textit{Design 1}: $\Sigma = \left((0.5)^{|j-k|}\right)_{jk}$;
\item[] \textit{Design 2}: Uncorrelated block structure: $$\Sigma = \left(\begin{array}{cc}
  \Sigma_{Z_1Z_1} & \Sigma_{Z_1X_{-1}}\\
  \Sigma_{Z_1X_{-1}}^T & \Sigma_{X_{-1}X_{-1}}\\
  \end{array}\right),$$
 where $\Sigma_{Z_1X_{-1}}$ is a $p_z\times p_w$ matrix of zeros,  $\Sigma_{Z_1Z_1}  = \left((0.5)^{|j-k|}\right)_{1\leq j,k\leq p_z}$,
  and  $\Sigma_{X_{-1}X_{-1}}  = \left((0.5)^{|j-k|}\right)_{1\leq j,k\leq p_w}$.
\end{itemize}

\begin{figure}[h!]
\captionsetup{format=plain}
   \hspace*{-.02\textwidth}
  \subfloat[\label{fig_Application_1_Overidentified_TP}{}][$n=p = q = 100$]{
      \includegraphics[width=0.37\linewidth]{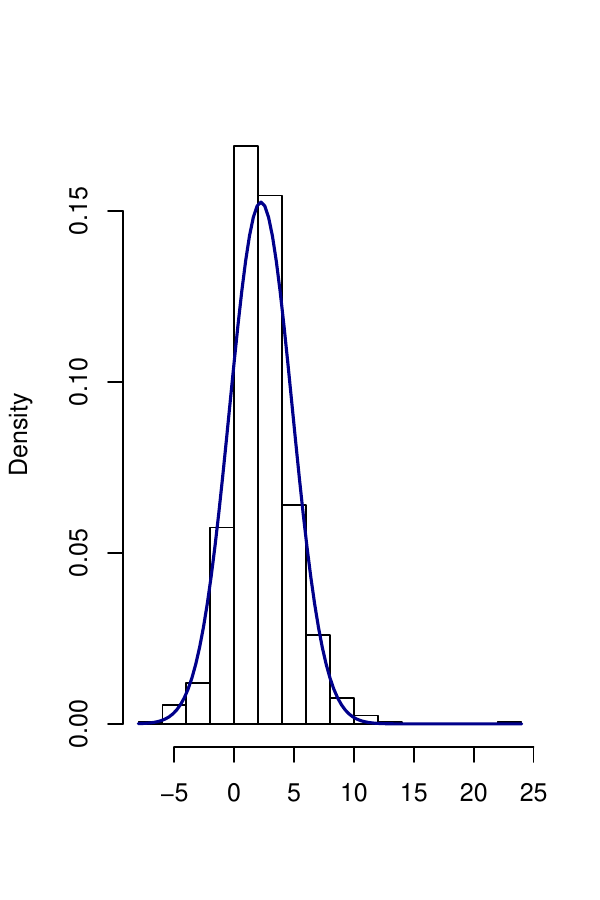}}
 \hspace*{-.045\textwidth}
  \subfloat[\label{fig_Application_1_Overidentified_H}{}][$p = q = 150$, $n=100$]{
      \includegraphics[width=0.37\linewidth]{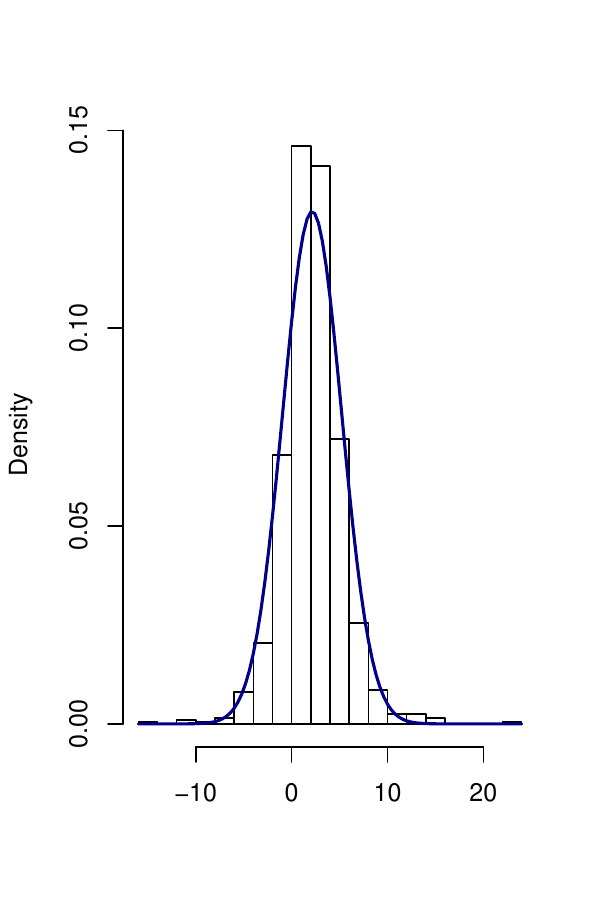}}
%  \hspace{0.1\linewidth}
 \hspace*{-.05\textwidth}
  \subfloat[\label{fig_Application_1_Overidentified_RM}{}][$p = q = 200$, $n=100$]{
      \includegraphics[width=0.37\linewidth]{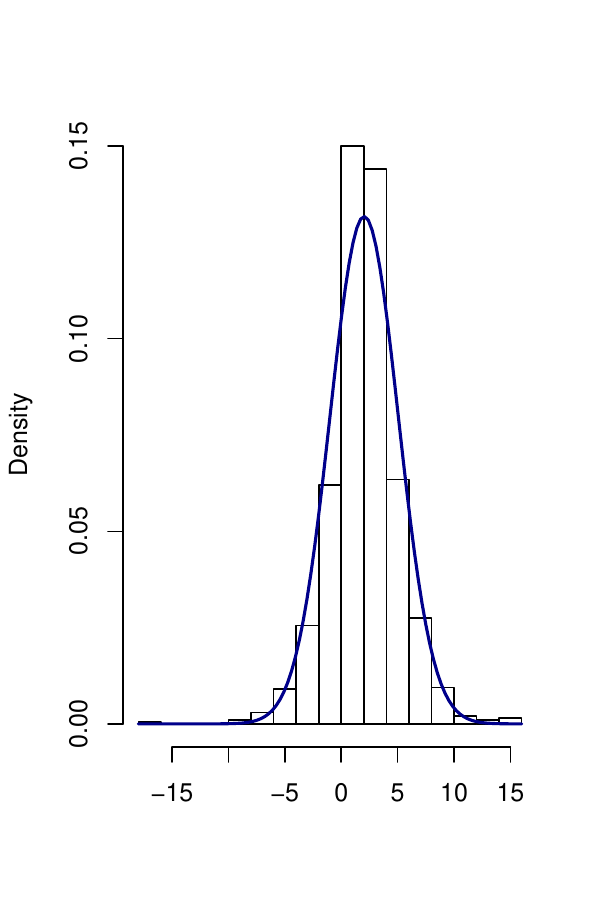}}
  \caption{{\small Heteroskedastic case - Design 2. Histograms approximating the sampling distribution of $\wh\beta_1$ for the simulation design \eqref{simulation:design:heteroscedasticity} based on a Monte Carlo experiment with $1000$ iterations.}}
  \label{fig_histogram:heteroskedasticity:2}
\end{figure}

In the rest of this section, we fix the degree of endogeneity and the strength of the instruments by setting $\rho = 0.5$ and $\alpha_1=1$. The other parameters are set as in the previous simulation with homoscedastic errors. The covariance estimator under heteroscedasticity is implemented as the matrix $\widehat\Omega$ in \eqref{Omega_hat_def}.\\
\indent In Tables \ref{Table:Bias:Exact_sparsity:heteroskedasticity:1} and \ref{Table:Bias:Exact_sparsity:heteroskedasticity:2} we report the absolute value of the estimated bias -- computed as the difference between the average over $1000$ Monte Carlo iterations and the true value of $\beta_1$ given by $2$ -- for our desparsified IV Lasso estimator $\wh\beta_1$ and for the IV Lasso estimator $\wtl\beta_1$. Table \ref{Table:Bias:Exact_sparsity:heteroskedasticity:1} refers to \textit{Design 1} while Table \ref{Table:Bias:Exact_sparsity:heteroskedasticity:2} refers to \textit{Design 2}. We see that the bias of $\wh\beta_1$ is again considerably smaller than the one of the initial IV Lasso estimator $\wtl\beta_1$ in absolute value. Compared to the bias reported in Table \ref{Table:Bias:Exact_sparsity}, in presence of heteroskedasticity the bias is larger in absolute value. However the bias of our estimator $\wh\beta_1$ is less affected by heteroscedasticity than the bias of the initial IV Lasso estimator $\wtl\beta_1$. In addition, we report the average coverage of our confidence interval for $\beta_1$ at the confidence level of $95\%$. We see that the coverage increases with  $q$. We also report the average coverage of the intervals for individual coefficients corresponding to variables in either $S_0$ or $S_0^c$. For the \textit{Design 2} we also outline in Table \ref{Table:Bias:Exact_sparsity:heteroskedasticity:2} the effect of augmenting $p$. %We notice that these coverages are slighter lower as $q=300$. One reason might be that for very large $q$ it is more difficult to identify the true active set.
Figure \ref{fig_histogram:heteroskedasticity:2} repots the histograms relative to \textit{Design 2} which show that the distribution of $\wh\beta_1$ is more and more concentrated around the true value of $\beta_1$ as $n$ and $q$ increase.
\begin{table}[ht!]
\captionsetup{format=plain}
\centering
\renewcommand{\arraystretch}{1.3}
  {\small \begin{tabular}{|c||c|c|c|c|c|}
    \hline
    & Absolute & Absolute & Coverage for & Coverage for & Coverage for\\
    & mean bias$(\wh\beta_1)$ & mean bias$(\wtl\beta_1)$ & $\beta_1$ & $S_0$-coefficients & $S_0^c$-coefficients\\
    \hline\hline
 $p=q=100$ & 0.326
& 1.605
& 0.908
& 0.880
& 0.967 \\
    \hline
 $p=q=150$ & 0.387
& 1.735
& 0.908
& 0.894
& 0.969\\
 \hline
 $p=q=200$ & 0.450
& 1.820
& 0.922
& 0.901
& 0.969\\
 \hline
  \end{tabular}}
  \caption{\footnotesize{Heteroskedastic case - Design 1 from model \eqref{simulation:design:heteroscedasticity} with $n = 100$  and varying $p$ and $q$. The same explanations as in Table \ref{Table:Bias:Exact_sparsity} apply.}} \label{Table:Bias:Exact_sparsity:heteroskedasticity:1}
\end{table}

\begin{table}[ht!]
\captionsetup{format=plain}
\centering
\renewcommand{\arraystretch}{1.3}
 {\small \begin{tabular}{|c||c|c|c|c|c|}
    \hline
    & Absolute & Absolute & Coverage for & Coverage for & Coverage for\\
    & mean bias$(\wh\beta_1)$ & mean bias$(\wtl\beta_1)$ & $\beta_1$ & $S_0$-coeff. & $S_0^c$-coeff.\\
    \hline\hline
 $p=q=100$ & 0.236
& 1.734
& 0.936
& 0.879
& 0.966\\
\hline
$p=100$, $q=150$ & 0.016
& 0.665
& 0.794
& 0.892
& 0.966\\
    \hline
 $p=q=150$ &0.138
& 1.800
& 0.936
& 0.894
& 0.969\\
 \hline
 $p=150$, $q=200$ &0.029
& 0.653
& 0.836
& 0.897
& 0.965\\
 \hline
 $p=q=200$ & 0.029
& 1.885
& 0.939
& 0.899
& 0.968\\
 \hline
  \end{tabular}}
  \caption{\footnotesize{Heteroskedastic case - Design 2 from model \eqref{simulation:design:heteroscedasticity} with $n = 100$  and varying $p$ and $q$. The same explanations as in Table \ref{Table:Bias:Exact_sparsity} apply.}} \label{Table:Bias:Exact_sparsity:heteroskedasticity:2}
\end{table}

%============================================================================================================================
\subsection{Increasing number of endogenous variables}\label{ss_Series_approximation}
This simulation corresponds to Example \ref{example:series:approximation} about series approximation. Let $\phi^J(X_1):= (\phi_1(X_1), \ldots, \phi_J(X_1))^T$ be a $J$-vector of basis functions. We generate i.i.d. data from the following model
\begin{align}
    Y & = \varphi(X_1) + \beta_{-1}^T X_{-1}+ U,\qquad X = (\phi^J(X_1), X_{-1}^T)^T,\nonumber\\
    X_1 & = \alpha_1^T Z_1 + \alpha_{-1}^T X_{-1}+ V/2,\qquad Z= (Z_1^T, X_{-1}^T)^T,\label{simulation:design:series:approximation}
  \end{align}
with $(U,V,Z)$ is generated from \eqref{gen:data}
again with $\Sigma = \left((0.5)^{|j-k|}\right)_{jk}$. Moreover, $\varphi(X_1) = X_1^2/4$ so that if $(\phi_j)_j$ are polynomials we have that $\phi^J(X_1) = (1, X_1, (X_1/2)^2, \ldots, (X_1/J)^J)$ and $\beta_1 = (0,0,1,0,\ldots,0)^T$.
We take $J=10$, $\dim(X_{-1}) = 100 - J$ and $q = 100$ exogenous variables (included and excluded covariates). The parameters are set in the following way: $\beta_{-1,j} = 1 + (j-1)*c$
for $1\leq j \leq 50$, where $c$ is a constant such that the parameters $\beta_{-1,j}$ are equispaced between $1$ and $3$, $\alpha_{1,j}$ has components equispaced between 1 and 0.5, $\beta_{-1,j} = 0$ for $51 \leq j \leq (p-J)$and $\alpha_{-1,j} = j^{-3}/2$ for $1\leq j\leq 50$. The results of this simulation are reported in Table \ref{Table:Bias:Exact_sparsity:series:approximation}. We report the absolute mean of the bias for the desparsified IV estimator and the initial IV Lasso estimator of the parameter $\beta_{1,3} = 1$, that is, the coefficient of the second order polynomial. We see that we obtain some undercoverage for the coefficient $\beta_{1,3}$ but the coverages for $S_0$-coefficients and $S_0^c$-coefficients are close or beyond the $95\%$ nominal level.
\begin{table}[ht!]
\captionsetup{format=plain}
\centering
\renewcommand{\arraystretch}{1.3}
  {\small \begin{tabular}{|c||c|c|c|c|c|}
    \hline
    & Absolute & Absolute & Coverage for & Coverage for & Coverage for\\
    & mean bias$(\wh\beta_{1,3})$ & mean bias$(\wtl\beta_{1,3})$ & $\beta_{1,3}$ & $S_0$-coefficients & $S_0^c$-coefficients\\
    \hline\hline
 $p=q=100$ & 0.070
& 0.890
& 0.900
& 0.967
& 0.966\\
 \hline
 $p=q=150$ & 0.033
&0.800
& 0.907
& 0.955
& 0.977\\
 \hline
 $p=q=200$ & 0.048
& 0.706
& 0.883
& 0.932
& 0.978\\
 \hline
  \end{tabular}}
  \caption{\footnotesize{The simulation design is \eqref{simulation:design:series:approximation} with $n = 100$. The same explanations as in Table \ref{Table:Bias:Exact_sparsity} apply. 
  }} \label{Table:Bias:Exact_sparsity:series:approximation}
\end{table}

%============================================================================================================================
\section{Application to the Logit Demand Estimation}\label{Application}
In this section we apply our method to estimate the price coefficient in a logit model of demand for automobiles using market share data. This application follows the empirical illustration in \cite{ChernozhukovHansenSpindler2015} and the aim is to estimate the price effect on the market share of a particular car. We consider the following system of equations:
\begin{eqnarray*}
  \log(s_{it}) - \log(s_{0t}) & = & \beta_{0}^{0} p_{it} + x_{it}^T\beta_{1}^{0} + u_{it},\\
  p_{it} & = & z_{it}^T\alpha_{0,0} + x_{it}^T\alpha_{0,1}  + \varepsilon_{it},
\end{eqnarray*}
where $s_{it}$ is the market share of product $i$ in market $t$, $s_{0t}$ denotes the outside option, $p_{it}$ is the price which is endogenous, $x_{it}$ are observed product characteristics which are exogenous, and $z_{it}$ is a set of instrumental variables.\\
\indent In our application we use the same product characteristics as in \cite{ChernozhukovHansenSpindler2015} and \cite{BerryLevinsohnPakes_1995}, that is, $x_{it}$ contains an air conditioning dummy, horsepower divided by weight, miles per dollar, vehicle size and a time trend. We center all the variables in order to eliminate the constant. The instruments for price are formed by using the idea developed in \cite{BerryLevinsohnPakes_1995} that characteristics of other products satisfy an exclusion restriction of the type $ \mathbb{E}[u_{it}|x_{jt'}] = 0$ for any $t'$ and any $j\neq i$. Therefore, any function of characteristics of other products may be used as an instrument for price. We follow \cite{ChernozhukovHansenSpindler2015} and form instruments as
\begin{equation}\label{eq:instruments:application}
  z_{k,it} = \left(\sum_{j\neq i, j\in\mathcal{I}_f}x_{k,jt}, \sum_{j\neq i, j\notin\mathcal{I}_f}x_{k,jt}\right),
\end{equation}
\noindent where $x_{k,it}$ and $z_{k,it}$ denote the $k$-th element of $x_{it}$ and $z_{it}$, respectively, and $\mathcal{I}_f$ denotes the set of products produced by firm $f$. In this way we have a set of $10$ excluded instruments.\\
\indent In addition, because economic theory does not specify the functional form in which the elements of $x_{it}$ enter the regression model, we also consider first-order interaction terms of the variables in $x_{it}$, and quadratic and cubic transformations of the continuous variables in $x_{it}$ for a total of $18$ new variables. In this way, we have a vector of augmented controls, denoted by $x_{it}^a$ that contains $x_{it}$ and these new variables. The corresponding vector of augmented excluded instruments is then given by $z_{it}^a$ where $z_{it}^{a}$ is constructed as in \eqref{eq:instruments:application} but with $x_{k,jt}$ replaced by $x_{k,jt}^a$. By using the generic notation in the paper: $X = (p_{it},x_{it}^{aT})^T$, $Z = (z_{it}^{aT},x_{it}^{aT})^T$, and $\beta^0 = (\beta_{0}^{0},\beta_1^{0T})^T$.\\
\indent In our data set, we have a total of $n = 2217$ observations, $23$ augmented controls, and $71$ augmented instruments $Z$.
%To measure the strength of the instruments we look at the mapping properties of the matrix $M$ in this application.
 According to our theory, the strength of identification is measured through the parameter $\omega$.
In the non-augmented framework with $10$ excluded instruments, the estimated $\omega$ is equal to $6.44\cdot10^{-06}$ and hence, relatively small.
%which suggests that we are in a framework with strong identification.
When we augment the number of controls and instruments the estimate of $\omega$ increases. This means that adding polynomial transformations and interactions, if on the one hand makes the model more flexible, on the other hand reduces the strength of identification, which is not surprising. This is not a problem since our approach is robust to semi strongly identified models.

In our application we estimate the covariance matrix to construct the confidence intervals by using our heteroscedastic robust estimator. In Table \ref{Table:logit:demand:estimation} we show the results obtained with different estimators. Together with the point estimate, we also report the lower and upper bound of the $95\%$-confidence interval. We first compute the OLS and 2SLS estimators obtained without augmenting the controls and the instruments. Then, we show the results obtained with augmented controls and instruments with three estimators: the OLS, the 2SLS and our desparsified IV Lasso estimator. To incorporate uncertainty induced by sample splitting for the selection of the tuning parameters, we use the finite-sample adjustments proposed by \cite{Chernozhukov2018EJ}. Specifically, we present estimation results as the median of desparsified IV Lasso estimate for 200 different sample splits.
%We use $200$ seeds selected to give results plausible from an economic theory point of view.
 Here, we retain estimates of our desparsified IV Lasso estimate gives a low number of products with inelastic demand. As we explained below, inelastic demand is unrealistic in a setup of firms maximizing their profit. The standard deviation is computed from the median, over the same $200$ seeds, of the adjusted variances (following the variance adjustment in equation (3.14) in \cite{Chernozhukov2018EJ}).

We see that the estimated price coefficient becomes larger in absolute value when we move from the Baseline OLS ($-0.0886$ with a standard deviation of $0.0043$) to the Baseline 2SLS ($-0.1419$ with a standard deviation of $0.0119$), which can be interpreted as the fact that the OLS estimator is biased because of endogeneity of price. The magnitude of the OLS estimated price coefficient increases when we use augmented controls (the Augmented OLS estimate is $-0.0991$ with a standard deviation of $0.0046$) and, when we use augmented controls and instruments, the Augmented 2SLS price coefficient estimate is -0.1273 with a standard deviation of $0.0076$. The largest value in absolute value is obtained with our desparsified IV Lasso estimator which gives a price coefficient estimate equal to $-0.2104$ with a (finite-sample adjusted) standard deviation of $0.0306$.
The estimated price coefficient that we obtain with our estimator is similar to the one in \cite{ChernozhukovHansenSpindler2015} based on the double-selection approach, which is equal to $-0.221$. The latter is contained in our $95\%$-confidence interval for $\beta_0^0$. The slight difference is mainly due to the randomness in choosing the tuning parameters.\\ 
\indent  Notice that as we move from the baseline results to the results based on augmented controls and instruments, the estimates become more plausible from an economic theory point of view. Indeed, in our setup firms maximizing their profit should face elastic demand for all products. In line with this insight, whereas the baseline OLS (resp. 2SLS) point estimates imply inelastic demand for 1502 (resp. $670$) products, our desparsified IV Lasso estimate inelastic demand for only $32$ products using augmented controls and variables. The number of products with inelastic demand are reported on the last column of Table \ref{Table:logit:demand:estimation}. For our desparsified IV estimator, the number of inelastic is larger than the one based on the double-selection procedure of \cite{ChernozhukovHansenSpindler2015} which is equal to $12$. This difference is due to the fact that our estimate of the price coefficient is slightly lower than the double-selection based estimate, as discussed above.\\% and have been calculated as the number of products for which $\wh\beta_{0}^0 p_{it}(1 - s_{it}) > -1$.
\begin{table}[ht!]
\captionsetup{format=plain}
\centering
\renewcommand{\arraystretch}{1.3}
  {\small \begin{tabular}{|c||c|c|c|c|}
    \hline
    & Price Coefficient & Lower & Upper & Number Inelastic\\
    \hline\hline
Baseline OLS & -0.0886 & -0.0971 & -0.0802 & 1502\\
Baseline 2SLS & -0.1419 & -0.1651 & -0.1186 & 670\\
Augmented OLS & -0.0991 & -0.1081 & -0.0901 & 1405\\
Augmented 2SLS & -0.1273 & -0.1423 & -0.1124 & 874\\
Desparsified IV & -0.2104 & -0.2704 & -0.1504 & 32\\
 \hline
  \end{tabular}}
  \caption{\footnotesize{Logit Demand Estimation. Comparison of different estimators for the price coefficient $\beta_0^0$. ``Baseline OLS'' refers to the OLS estimate obtained with the non augmented controls $x_{it}$, ``Augmented OLS'' refers to the OLS estimate obtained with the augmented controls $x_{it}^a$, ``Baseline 2SLS'' refers to the 2SLS estimate obtained with the non augmented controls $x_{it}$  and the non augmented instruments $z_{it}$, ``Augmented 2SLS'' refers to the 2SLS estimate obtained with the augmented controls $x_{it}^a$ and the augmented instruments $z_{it}^a$, ``Desparsified IV'' refers to our desparsified IV Lasso estimate obtained with the augmented controls $x_{it}^a$ and the augmented instruments $z_{it}^a$. ``Lower'' and ``Upper'' denote the lower and upper bound of the $95\%$-confidence interval for $\beta_0^0$. Finally, ``Number Inelastic'' refers to the point estimate of the number of products for which demand is estimated to be inelastic.}} \label{Table:logit:demand:estimation}
\end{table}
\indent Overall, we see that our desparsified IV estimator and inference procedure perform well in empirical applications and give plausible results. In addition, because our procedure is robust to heteroscedasticity, our inference remains valid when the regression error term is  heteroscedastic. 

\appendix

\section{Appendix: Proofs}\label{app:proofs}
Let Assumption \ref{A:Z} hold. By using the Cauchy-Schwarz inequality, the definition of $s_M$, and the assumption that $\lambda_{\max}(\Sigma)=O(1)$ and $\lambda_{\max}(M^T\Theta M)=O(1)$ we obtain
\begin{align*}
\|M\|_1&\leq \sqrt{s_M}\max_{1\leq j\leq p}\|M_j\|_2\\
&=O\big(\sqrt{s_M}\max_{1\leq j\leq p}\|\Theta^{1/2} M e_j\|_2\big)\\
&=O\big(\sqrt{s_M}\big)
\end{align*}
where $M_j$ denotes the $j$--th column of the matrix $M$. Similarly, the sparsity constraint on $\Theta$ implies
\begin{align*}
\|\Theta\|_1\leq \sqrt{s_{\max}}\max_{1\leq j\leq q} \|\Theta_j\|_2 = O(\sqrt{s_{\max}}).
\end{align*}

For the next proofs, we require the following notation.
 For $j=1,\dots,p$, recall the definition $\gamma_j := \argmin_{\gamma\in\mathbb{R}^{p-1}}\|(\Theta^{1/2}M)_j - (\Theta^{1/2} M)_{-j}\gamma\|_2^2$. We also define $\tau_j^2:=\|( \Theta^{1/2} M)_j - (\Theta^{1/2} M)_{-j}\gamma_j\|^2_2$. Introduce a vector $\Gamma_j:=(\Gamma_{kj})_{k=1}^p$ with $\Gamma_{kj}=-\gamma_{kj}$ for $k\neq j$ and otherwise $1$,
where $\gamma_{kj}$ is the $k$--th entry of $\gamma_j$. Then, we have $\tau_j^{2}=\Gamma_j^T  M^T \Theta M \Gamma_j$ since $\Theta^{1/2} M\Gamma_j = (\Theta^{1/2} M)_j - (\Theta^{1/2} M)_{-j}\gamma_j$. It also holds $\tau_j^2=1/\Theta^M_{jj}$, which can be seen as follows.
The first order condition for $\gamma_j$ yields
\begin{align*}
 (\Theta^{1/2}M)_{-j}^T \Theta^{1/2}M \Gamma_j =0,
\end{align*}
and thus,
\begin{align*}
 M^T\Theta M \Gamma_j
   =  \left((\Theta^{1/2} M)_j^T\Theta^{1/2} M \Gamma_j\right)e_j=  \Gamma_j^T  M^T \Theta M \Gamma_j e_j = \tau_j^{2} e_j,
\end{align*}
\noindent where we have used the fact that $\Theta^{1/2} M\Gamma_j = (\Theta^{1/2} M)_j - (\Theta^{1/2} M)_{-j}\gamma_j$ together with the first order condition for $\gamma_j$ to get the second equality.
Further, by premultiplying with $\Theta^M$ we obtain
\begin{align*}
 \Gamma_j = \tau_j^{2}\, \Theta^M e_j
\end{align*}
and since $e_j^T\Gamma_j=1$ we obtain $\tau_j^2=1/\Theta^M_{jj}$.
By the definition of $\omega $ we obtain the following lower bound for $\tau_j$:
\begin{align}\label{lower:bound:tau}
\tau_j^2 = 1 / \Theta_{jj}^M \geq 1 / \lambda_{\max}(\Theta^M)= \lambda_{\min}(M^T \Theta M) =\omega ^{-1},
\end{align}
 which we will use in the following proofs.
 Reversely, $\tau_j$ is bounded from above by the maximal eigenvalue of $M^T \Theta M$ which we assume to be bounded. This implies that
\begin{align*}
\|\gamma_j\|_1^2 & \leq C \Big(s_j^M\|\gamma_j\|_2^2+(\log(q))^2/n\Big)\\
&\leq C \Big(s_j^M+(\lambda_j^M)^2\Big).
\end{align*}
where we have used the upper bound $\|\gamma_j\|_1 \leq \|\gamma_{j,S_j}\|_1 + \|\gamma_{j,S_j^c}\|_1$, the Cauchy-Schwarz inequality and \eqref{eq:gamma:s:j} to get the first inequality.
Below we also use for matrices $A$ and $B$ the inequalities
\begin{align*}
 \|A B \|_\infty \leq \|A  \|_\infty \| B \|_1 \qquad \mbox { and  }  \qquad \|A B \|_\infty \leq \|B  \|_\infty \| A^T \|_1.
 \end{align*}

\subsection{Proofs of the Main Results}
\begin{proof}[\textsc{ Proof of Lemma \ref{lem:comp}.}]
%By using the inequality $\|\Sigma^{-1/2}v\|_2\leq \|v\|_2/\sqrt{\lambda_{\min}(\Sigma)}$ for all
We make use of the inequality
\begin{align*}
\|M^T\Sigma^{-1} v\|_2\leq \|v\|_2\sqrt{\lambda_{\max}(MM^T)}/\lambda_{\min}(\Sigma)
\end{align*}
for all
$v\in\mathbb{R}^q$ and the fact that $\Sigma$ has eigenvalues uniformly bounded away from zero by Assumption \ref{A:Z} \textit{(iii)}. Consequently, sub-Gaussianity of $Z$ implies sub-Gaussianity of $\wtl{Z} := M^T\Sigma^{-1} Z$. We make use Lemma 5.2 (and the proof of Theorem 2.4) in \cite{vandegeer2014} to the reduced form model
\begin{align*}
Y=\widetilde Z^T\beta^*+V,
\end{align*}
where $\beta^*:=\Sigma^{-1/2}M\beta^0$ and $V=Y - Z^T\Sigma^{-1} \EE[YZ]$. Hence, sub-Gaussianity of $\widetilde Z$ and Assumption \ref{A:Z} \textit{(iii)} imply
%\begin{align}\label{comp:cond}
%\|(\Sigma^{-1/2} M\beta)_{S_0}\|_1^2\leq C s_0\, \beta^TM^T\Sigma^{-1}\widehat\Sigma \Sigma^{-1} M\beta
%\end{align}
\begin{align}\label{comp:cond}
\|\beta_{S_0}\|_1^2\leq C s_0\, \beta^TM^T\Sigma^{-1}\widehat\Sigma \Sigma^{-1} M\beta
\end{align}
wpa1, for  all $\beta\in\mathcal B$.
%In addition, from Assumption \ref{A:IV} and the definition of $\omega_1$ we infer
%\begin{align*}
%\cB := \set{ \|\beta_{\widetilde S_0^c}\|_1\leq 3\|\beta_{\widetilde S_0}\|_1}\subset \set{\|(\Sigma^{-1/2} M\beta)_{S_0^c}\|_1\leq 3\eta^{-1/2}\|(\Sigma^{-1/2} M\beta)_{S_0}\|_1}.
%\end{align*}
%Consequently, \eqref{comp:cond} holds on $\cB$ and then by the definition of $\omega_1$, inequality \eqref{comp:cond:struc} follows for all $\beta\in\cB$.
\end{proof}

%==============================================================================
%==============================================================================
\begin{proof}[\textsc{ Proof of Theorem \ref{theo:main}.}]
  The proof is based on the decomposition
   \begin{align*}
  \Delta = \sqrt{n}\big(\wh\Theta^M\wh M^T\wh\Theta \wh M - I_p\big)(\wtl\beta - \beta^0) - \sqrt{n}\,\wh{\Theta}^M \wh M^T \wh\Theta(\wh M - \widetilde{M})(\wtl\beta - \beta^0).
  \end{align*}
 We observe
\begin{align*}
\|\Delta\|_\infty/\sqrt n&\leq\|\big(\wh\Theta^M\wh M^T\wh\Theta \wh M - I_p\big)(\wtl\beta - \beta^0)\|_\infty+\|\wh{\Theta}^M \wh M^T \wh\Theta(\wh M - \widetilde{M})(\wtl\beta - \beta^0)\|_\infty\\
&\leq\|\wh\Theta^M\wh M^T\wh\Theta \wh M - I_p\|_\infty\|\wtl\beta - \beta^0\|_1
+\|\wh{\Theta}^M \wh M^T \wh\Theta\|_{op,\infty}\|(\wh M - \widetilde{M})(\wtl\beta - \beta^0)\|_\infty.
%&\leq \Big(\|\wh\Theta^M\wh M^T\wh\Theta \wh M - I_p\|_\infty\Big)\|\wh\beta - \beta^0\|_1\\
\end{align*}
Further, the upper bound given in \eqref{bound:approx} implies that
\begin{align*}
 \|\Delta\|_\infty&\leq\sqrt n\max_{1\leq j\leq p}\set{\lambda_j^M/\,\widetilde\tau_j^2}\, \|\widetilde{\beta} - \beta^0\|_1+\sqrt n\|\wh\Theta^T\wh M(\wh{\Theta}^M)^T\|_1 \|\wh M - \widetilde{M}\|_\infty\|\wtl\beta - \beta^0\|_1.
\end{align*}
By the definition of the regularized estimator $\widehat M$ given in \eqref{eq_M_est_thresholding} it holds for all $j,k$:
\begin{align*}
|\wtl M_{jk}-\wh M_{jk}|& = |\wtl M_{jk}| \1\set{ |\wtl M_{jk}| < C_0\sqrt{\log(q)/n}  }\\
&<C_0\sqrt{\log(q)/n },
\end{align*}
which implies
\begin{align}\label{ineq:M}
\|\wtl M-\wh M\|_\infty < C_0 \sqrt{\log(q)/n} .
\end{align}
Thus, using that $\lambda_j^M\sim \log (q)/\sqrt{ n}$ uniformly in $j$, by Assumption \ref{A:M} (i) we obtain
\begin{align*}
 \|\Delta\|_\infty&\leq C\log(q)\,\Big(\max_{1\leq j\leq p}\,\widetilde\tau_j^{-2}+\|\wh\Theta^T\wh M(\wh{\Theta}^M)^T\|_1\Big)\|\widetilde{\beta} - \beta^0\|_1.
\end{align*}
In the following, we consider the events
   \begin{align*}
   \mathcal C := \set{  \|\beta_{S_0}\|_1^2\leq s_0\beta^T \wh M^T(\wh \Theta+\wh \Theta^T) \wh M\beta/c^2\text{ for all }\|\beta_{S_0^c}\|_1\leq 3\|\beta_{S_0}\|_1}
   \end{align*}
and $\cT:=\set{\|\wh M^T\, (\wh \Theta+\wh \Theta^T) \mathbf Z^T \mathbf U/n+\wh M^T(\wh \Theta+\wh \Theta^T) (\wtl M - \wh M )\beta^0\|_\infty\leq C \lambda^*}$ for some sufficiently large constant $C$ where $\lambda^*>C c_1 \lambda$ for some $c_1>1$ and recall $\lambda\sim \log(q)/\sqrt{n}$.

On the event $\mathcal C \cap \cT$ we have
 \begin{align*}
    \|\wtl\beta - \beta^0\|_1\leq C s_0 \log (q)/\sqrt{n},
    \end{align*}
which follows directly from \cite[Theorem 2.2]{van2016}.\footnote{Apply \cite[Theorem 2.2]{van2016} with, in their notation, $L=3$, $\wh \phi^2(3, \wtl S_0)= c^2$, $X= (\wh \Theta+\wh \Theta^T)^{1/2} \wh M$, $Y= (\wh \Theta+\wh \Theta^T) ^{1/2} \mathbf Z^T \mathbf Y$, $\epsilon = (\wh \Theta+\wh \Theta^T)^{1/2} (\mathbf Z^T \mathbf Y -  \wh M \beta^0)$.}
    From the proof of Proposition \ref{prop:main} in Appendix \ref{Appendix:A:2} we also have that $\wtl\tau_j^2$ is a consistent estimator of $\tau_j^2$. Further, Propositions \ref{theo:bound} and  \ref{prop:main} together with the  lower bound \eqref{lower:bound:tau} yield
    \begin{align*}
\|\Delta\|_\infty\1_{\mathcal C \cap \cT}=O_p\Big(s_0\log(q)^2/\sqrt{n}\,\max\big(\omega , \|\Theta M{\Theta}^M\|_1\big)\Big).
\end{align*}
It is thus sufficient to show $\1_{\mathcal C \cap \cT}=1$ wpa1. We proceed in two steps and control the sets $\cT$ and $\mathcal C$ separately.
To handle the set $\cT$ note that
\begin{align*}
\big\|\wh M^T\, \wh \Theta \mathbf Z^T \mathbf U/n&+\wh M^T\wh\Theta (\wtl M - \wh M )\beta^0\big\|_\infty\\
&\leq \underbrace{\|\mathbf U^T\mathbf Z\,\Theta\,M/n\|_\infty}_{I}
+\underbrace{\|( \wh M^T\wh \Theta - M^T\Theta)\big(\mathbf U^T\mathbf Z/n+(\wtl M - \wh M)\beta^0\big)\|_\infty}_{II}\\
&\qquad\qquad+\underbrace{\|M^T\Theta(\wtl M - \wh M)\beta^0\|_\infty}_{III}.
\end{align*}
To bound $I$,  we make use of Nemirovski's inequality (see, for instance, p. 509 in \cite{buhlmann2011}) and $\EE[U^2|Z]\leq \sigma^2$ to get
\begin{align*}
\EE\Big(\max_{1\leq j\leq p}\big| (\mathbf U^T\mathbf Z\,\Theta\, M)_j/n\big|\Big)^2
&\leq 8\log(2p)\frac{1}{n^2}\sum_{i=1}^n \EE \max_{1\leq j\leq p}| U_i( Z_i^T\,\Theta\, M)_{j}|^2\\
&\leq 8\log(2p)n^{-1}  \sigma^2 \EE \max_{1\leq j\leq p}|( Z^T\,\Theta\, M)_{j}|^2
\end{align*}
and hence, we obtain $ I=O_p\big(\sqrt{\EE \|Z^T\,\Theta\, M\|_\infty^2\log(p)/ n}\big) = O_p\left(\log(p)/\sqrt{n}\right)$ by using Assumption \ref{A:M} $(ii)$. Under Assumption \ref{A:M} $(i)$ we have that $\lambda\sim \log(q)/\sqrt{n}$ and thus, $I=O_p(\lambda)$.

Next, we consider $II$. We have
\begin{align*}
II
%&=\|( \wh \Theta\,\wh M - \Theta\,M)^T(\mathbf Z^T\mathbf U/n+\wtl M\beta^0 - \wh M\beta^0)\|_{\infty}\\
%&\leq\|( \wh \Theta\,\wh M - \Theta\,M)^T\|_{op,\infty} \|\mathbf Z^T\mathbf U/n+\wtl M\beta^0 - \wh M\beta^0\|_{\infty}\\
&= \|\wh M^T\wh \Theta - M^T\Theta\|_{op,\infty}\|\mathbf U^T\mathbf Z/n+(\wtl M - \wh M)\beta^0\|_\infty \\
&\leq \Big(\|\wh \Theta\|_{op,\infty}\|\wh M - M\|_1+\|M\|_1\|\wh \Theta - \Theta\|_{op,\infty}\Big)\big(\|\mathbf U^T\mathbf Z/n\|_\infty +\|\wtl M - \wh M\|_\infty\|\beta^0\|_1\big).
\end{align*}
Again, due to Nemirovski's inequality, we have $\|\mathbf U^T\mathbf Z/n\|_\infty =O_p(\sqrt{\log (q)/ n})$ under Assumption \ref{A:Z} $(iii)$ and condition $\EE[U^2|Z]\leq \sigma^2$ imposed in Assumption \ref{A:M} $(ii)$. Furthermore, $\|\wtl M - \wh M\|_\infty=O\big(\sqrt{\log(q)/n}\big)$ by inequality \eqref{ineq:M}. We also have $\|\wh M - M\|_1 =O_p\big(s_M \sqrt{\log(q)/n}\big)$ and $\|\wh\Theta-\Theta\|_{op,\infty}=O_p\big( s_{\max}\,\sqrt{\log(q)/n}\big)$ from Propositions \ref{th:norm1:M} and \ref{prop:main}.
Now using that $\|\beta^0\|_1=O(s_0)$ (since $\|\beta^0\|_1 \leq \|\beta_{S_0}^0\|_1 + C s_0\sqrt{\log(p)/n} \leq s_0 \|\beta^0\|_{\infty} + o(1)$ by using the fact that $\beta_{S_0}^0\in\mathcal{B}$, \eqref{approx:sparse:ineq} and \eqref{spars:rate:cond}, and $\|\beta^0\|_\infty=O(1)$) we obtain
\begin{align*}
 II&=O_p\big((1+\|\beta^0\|_1)\big(s_M\|\Theta\|_1+s_{\max}\|M\|_1\big)\log (q)/ n\big)\\
 &=O_p\big(s_0\max\big(s_M\sqrt{s_{\max}},s_{\max}\sqrt{s_M}\big)\log (q)/ n\big)\\
 &=o_p(\sqrt{\log(p)/n})
\end{align*}
employing \eqref{spars:rate:cond} in Assumption \ref{A:M} to get the last equality. Remark that to get the first equality we have used the fact that $\|\wh \Theta\|_{op,\infty} \leq \|\wh \Theta - \Theta\|_{op,\infty} + \|\Theta\|_1$ because $\Theta$ is symmetric, and by Proposition \ref{prop:main} $\|\wh\Theta - \Theta\|_{op,\infty} = O_p(s_{\max}\,\sqrt{\log(q)/n})$ which is negligible with respect to the other terms under \eqref{spars:rate:cond}.
Consider $III$.
We have
\begin{align}\label{proof_term_III}
  III\leq \max_j\big|(\Theta M)_j^T (\wh{M} - M)\beta^0\big|+\max_j\big|(\Theta M)_j^T (\wtl{M} - M)\beta^0\big|,
\end{align}
where the second summand can be bounded again by using Nemirovski's inequality:
\begin{align*}
\EE \big\|M^T\Theta (\wtl{M} - M)\beta^0\big\|_\infty^2&
=\EE \max_{1\leq j\leq p} \big|n^{-1}\sum_i (\Theta M)_j^T Z_i X_i^T\beta^0 - (\Theta M)_j^T M\beta^0\big|^2\\
&\leq 8\log(2p)n^{-1}  \EE \max_{1\leq j\leq p}|(\Theta M)_j^T Z X^T\beta^0|^2\\
&\leq 8\log(2p)n^{-1}  \EE \big[(X^T\beta^0)^2\|M^T\Theta Z\|_\infty^2\big]\\
&=O\big(\log(p)^2/n\big),
\end{align*}
where we have used Assumption \ref{A:M} $(ii)$ to get the last line.
 For the first summand on the right hand side of \eqref{proof_term_III} we observe
\begin{align}
\max_j\big|(\Theta M)_j^T (\wh{M} - M)\beta^0\big|&\leq \|\Theta M\|_{\infty} \|\wh M - M\|_1 \|\beta^0\|_1\nonumber\\
& = O_p\left( s_0\sqrt{s_{\max}}s_M \sqrt{\log(q)/n}\right)\nonumber\\
& = O_p\left(\log(q)/\sqrt{n}\right)\label{upper:bound:III}
\end{align}
due to Assumption \ref{A:Z} \textit{(iii)} which implies $\|\Theta\|_1 = O(\sqrt{s_{\max}})$, Assumption \ref{A:M} \textit{(ii)}, the second result of Proposition \ref{th:norm1:M} and
the first rate restriction imposed in Assumption \ref{A:M} \textit{(iii)}.

It remains to control ${\mathcal C}$.
By Lemma \ref{comp:cond:struc} it holds for all $\|\beta_{S_0^c}\|_1\leq 3\|\beta_{S_0}\|_1$ that
\begin{align*}
  \|\beta_{\wtl S_0}\|_1^2\leq s_0\, \beta^TM^T\Sigma^{-1}\widehat\Sigma \Sigma^{-1} M\beta/\wtl c^2
\end{align*}
wpa1, for some constant $\wtl c >0$. Thus, in order to prove that $\mathcal{C}$ holds wpa1 it suffices to show that for some sufficiently small constant $c^*>0$ it holds
\begin{align}\label{ineq:proof:1st}
s_0 \| M^T \Theta (\wh \Sigma- \Sigma ) \Theta M  \|_\infty \leq c^*/2\quad \text{wpa1 }
\end{align}
and
\begin{align}\label{ineq:proof:2nd}
 s_0 \| M^T \Theta  M -  \wh M^T\wh  \Theta\wh  M \|_\infty \leq c^*/2\quad \text{wpa1. }
  \end{align}
To prove \eqref{ineq:proof:1st}, note that $\| \wh \Sigma- \Sigma  \|_\infty \leq c^\prime \sqrt {{\log (q) }/{n}}$ wpa1 for some  constant $c^\prime >0$, see \textit{e.g.} \cite[Problem 14.2]{van2016}, and thus the result follows by
\begin{align*}
 s_0  \|  \Theta M  \|_1 ^2 \sqrt {\frac {\log (q) }{n}} \leq c^{**}
 \end{align*}
for some constant $c^{**}$ that is chosen small enough. This inequality is indeed satisfied due to $\|  \Theta M  \|_1 ^2\leq s_{\max}s_M$ and the rate requirement imposed in Assumption \ref{A:M} $(iii)$.

To show \eqref{ineq:proof:2nd} we first make the decomposition $\| M^T \Theta  M -  \wh M^T\wh  \Theta\wh  M \|_\infty  \leq \| M^T \Theta  M -  \wh M^T \Theta\wh  M \|_\infty + \| \wh M^T(\wh  \Theta - \Theta)\wh  M \|_\infty $. Then,
\begin{align*}
 s_0  \|  \wh M  \|_1 ^2 \| \wh  \Theta - \Theta  \|_\infty &\leq 2 s_0 \left(\|M\|_1^2 + \|\wh M - M\|_1^2\right) \|\wh \Theta - \Theta\|_{\infty}\\
 &\leq C s_0 s_M^2\big(1 + \log(q)/n\big) \sqrt{s_{\max}}\sqrt{\log(q)/n},%c^{**}/2
\end{align*}
wpa1,
where we have used Assumption \ref{A:M} $(ii)$ to get $\|M\|_1 \leq s_M\|M\|_{\infty} = O(s_M)$, the second result of Proposition \ref{th:norm1:M} and the result $\|\wh \Theta - \Theta\|_\infty =O_p (\sqrt{s_{\max}\log(q)/n})$ (see \cite{vandegeer2014}). Moreover,
\begin{align*}
  \| M^T \Theta  M &-  \wh M^T \Theta\wh  M \|_\infty \\
  &\leq \|M - \wh M\|_1 \|\Theta\|_1 \|M\|_{\infty} + \left(\|M\|_{\infty} + \|\wh M - M\|_1\right)\|\Theta\|_1 \|\wh M - M\|_1\\
  &\leq C s_M\sqrt{\log(q)/n}\sqrt{s_{\max}}\left(1 + s_M\sqrt{\log(q)/n}\right)
\end{align*}
wpa1,
where we have used Assumptions \ref{A:Z} $(iii)$ and \ref{A:M} $(ii)$ and the second result of Proposition \ref{th:norm1:M}.
Consequently, by the rate restriction $s_0  s_M\sqrt{s_{\max}}=o(\sqrt{n/\log(q)})$
in Assumption \ref{A:M} $(iii)$, result \eqref{ineq:proof:2nd} holds wpa1.
\end{proof}

\begin{proof}[\textsc{ Proof of Theorem \ref{th:comp:inf}.}]
We proceed in two steps. First, we show $\sqrt{n/(a^T\Omega\, a)}\,a^T\big(\wh \beta-\beta^0\big)\overset{d}{\rightarrow} \mathcal{N}(0,1)$.
 We make use of the decomposition
\begin{align*}
\sqrt{n/(a^T\Omega\, a)}\,a^T\big(\wh \beta-\beta^0\big)=&\underbrace{a^T\Theta^M M^T \Theta  \mathbf Z^T\mathbf U/\sqrt{n(a^T\Omega\, a)}}_{=I}\\
&+\underbrace{a^T\big(\wh\Theta^M \wh M^T\wh \Theta -\Theta^M M^T\Theta\big)\mathbf Z^T\mathbf U/\sqrt{n(a^T\Omega\, a)}}_{=II}\\
&+\Delta\|a\|_1/\sqrt{a^T\Omega\, a}.
\end{align*}
Since $\sqrt{a^T\Omega\, a}\geq \underline{\sigma} \sqrt{\omega } \|a\|_2$ it holds $\|a\|_1/\sqrt{a^T\Omega\, a}=O(\sqrt\omega)$ for all $a\in\mathcal A$. By Theorem \ref{theo:main} and rate condition \eqref{coro:cond}  we obtain $\Delta \|a\|_1/\sqrt{a^T\Omega\, a}=o_p(1)$. 
We have that $I\overset{d}{\rightarrow} \mathcal{N}(0,1)$ and moreover, $II=o_p(1)$ which can be seen as follows.
We observe
\begin{align*}
II\leq
\sqrt{\omega /(a^T\Omega\, a)}\, \|a\|_1\Big(&\|\wh\Theta^M -\Theta^M \|_{op,\infty}  \| M^T\Theta\mathbf Z^T\mathbf U\|_\infty/\sqrt{\omega n}\\
&+\|\wh M-M\|_1\|\Theta^M\|_1\|\wh \Theta\|_{op,\infty}\| \mathbf Z^T\mathbf U\|_\infty/\sqrt{\omega n}\\
&+\|\wh \Theta-\Theta\|_{op,\infty}\|M^T\Theta^M\|_1\| \mathbf Z^T\mathbf U\|_\infty/\sqrt{\omega n}\Big).
\end{align*}
Using Nemirovski's inequality as in proof of Theorem \ref{theo:main}  we have $\| M^T\Theta\mathbf Z^T\mathbf U\|_\infty/\sqrt n=O_p(\sqrt{\log(q)})$ and
$\|\mathbf Z^T\mathbf U\|_\infty/\sqrt n=O_p(\sqrt{\log(q)})$. Further, from $\sqrt{\omega /(a^T\Omega\, a)}\leq \underline\sigma^{-1} \|a\|_2^{-1} $ we infer
\begin{align*}
II= O_p\Big(\frac{\log(q)}{\sqrt n}\frac{\|a\|_1}{\|a\|_2}\big(\omega^{3/2} s_{\max}^M\sqrt{\log(q)}+\max\big(s_M\sqrt{s_{\max}},s_{\max}\sqrt{s_M}\big)\|\Theta^M\|_1/\sqrt \omega \big)\Big)
\end{align*}
using $\|\Theta M\|_1\leq \sqrt{s_Ms_{\max}}$.
The rate requirement imposed on $q$ implies the result.

Second, we establish consistency of covariance matrix estimation.
For the covariance matrix estimator $\widehat\Omega$ we conclude
\begin{align*}
\Big|\frac{a^T\wh\Omega\, a}{a^T\Omega\, a}-1\Big|&\leq
(a^T\Omega \, a)^{-1}\|a\|_1^2\|\wh\Omega-\Omega\|_\infty\\
&\leq
\underbrace{\big\|\wh\Theta^M \wh M^T\wh \Theta \big\|_1^2
\|n^{-1}\mathbf Z^T \text{diag}(\wh{ \mathbf U})^2\mathbf Z-\EE[U^2ZZ^T]\|_\infty}_{=A_1}\\
 &+\underbrace{\big\|\wh\Theta^M \wh M^T\wh \Theta -\Theta^M M^T\Theta\big\|_1\|\Theta^M M^T\Theta\|_1\|\EE[U^2ZZ^T]\|_\infty}_{=A_2}\\
 &+\underbrace{\big\|\wh\Theta^M \wh M^T\wh \Theta -\Theta^M M^T\Theta\big\|_1^2\|\EE[U^2ZZ^T]\|_\infty}_{=A_3}.
\end{align*}
Using again Nemirovski's inequality and $\EE[U^2|Z]\leq \sigma^2$ we obtain
\begin{align*}
\|n^{-1}&\mathbf Z^T \text{diag}(\wh{ \mathbf U})^2\mathbf Z-\EE[U^2ZZ^T]\|_\infty\\
&=\Big\|n^{-1}\sum_i \big(U_i+X_i^T(\beta^0-\wtl\beta)\big)^2Z_iZ_i^T-\EE[U^2ZZ^T]\Big\|_\infty\\
&\leq \Big\|n^{-1}\sum_i U_iZ_iZ_i^T-\EE[U^2ZZ^T]\Big\|_\infty
+ 2\Big\|(\beta^0-\wtl\beta)^Tn^{-1}\sum_i U_iX_i Z_iZ_i^T\Big\|_\infty\\
&+\Big\|n^{-1}\sum_i \big(X_i^T(\beta^0-\wtl\beta)\big)^2Z_iZ_i^T\Big\|_\infty\\
&\leq O_p\Big(\sqrt{\log(q)/n}\Big)+\|\beta_0-\wtl\beta\|_1\times O_p\big(\EE\|X\|_\infty^2\EE\max_{1\leq j,l\leq q}|Z_jZ_l|^2\big)\\
&\quad +\|\beta_0-\wtl\beta\|_1^2\times O_p\big(\EE\max_{1\leq j,l\leq p}|X_jX_l|^2\EE\max_{1\leq j,l\leq q}|Z_jZ_l|^2\big).
\end{align*}
%It is sufficient to bound.
Now using $ \|\wtl\beta - \beta^0\|_1=O_p\big(s_0 \sqrt{\log (p)/n}\big)$ we obtain the $A_1=o_p(1)$. Finally, by using a similar decomposition as for the bound of $II$, it is easy to see that $A_2=o_p(1)$ which implies $A_3=o_p(1)$.
\end{proof}

\subsection{Proofs of Bounds on Random Matrices}\label{Appendix:A:2}
\begin{proof}[\textsc{ Proof of Lemma \ref{theo:bound}.}]
The proof of \eqref{ineq:kkt} is given in \cite{vandegeer2014}. For completeness we provide the following arguments.
The KKT condition for $\wh\xi_{j}$ implies $\wh\tau_j^2 = \mathbf Z_j^T(\mathbf Z_j - \mathbf{Z}_{-j}\wh\xi_j)/n$.
Consequently, it holds $\mathbf Z_j^T \mathbf{Z}\wh\Theta_j/n = \mathbf Z_j^T(\mathbf Z_j - \mathbf{Z}_{-j}\wh\xi_j)/(n\wh\tau_j^2) = 1$. The KKT conditions also imply
 $\|\mathbf{Z}_{-j}^T\mathbf{Z}\wh\Theta_j\|_{\infty}/n \leq \lambda_j^{\Theta}/\wh\tau_j^2$ or
\begin{align*}
\big\|\wh\Sigma\wh \Theta_j - e_j \big\|_{\infty} \leq \lambda_j^{\Theta}/\wh\tau_j^2,
\end{align*}
where $e_j$ is the $j$--th unit column vector.

Proof of \eqref{bound:approx}. The KKT conditions for the nodewise Lasso \eqref{eq:gamma:M:Lasso} implies
\begin{align}
\wtl\tau_j^{2}&=\Big((\wh \Theta^{1/2}\wh M)_j - (\wh \Theta^{1/2}\wh M)_{-j}\wtl\gamma_j\Big)^T\Big((\wh \Theta^{1/2}\wh M)_j - (\wh \Theta^{1/2}\wh M)_{-j}\wtl\gamma_j\Big) + \lambda_j^M\|\wtl\gamma_{j}\|_1 \nonumber\\
&=\Big((\wh \Theta^{1/2}\wh M)_j - (\wh \Theta^{1/2}\wh M)_{-j}\wtl\gamma_j\Big)^T
(\wh \Theta^{1/2}\wh M)_j+\lambda_j^M\underbrace{\big(\|\wtl\gamma_{j}\|_1-\wtl\gamma_{j}^T\text{sign}(\wtl\gamma_{j})\big)}_{=0}\nonumber\\
&= \big(\wh \Theta^{1/2}\wh M \,\wtl \Gamma_j\big)^T
(\wh \Theta^{1/2}\wh M)_j.\label{Proof:Theorem:4:1:eq:1}
\end{align}
Consequently,  for all $1\leq j\leq p$:
\begin{align*}
 (\wh \Theta^{1/2}\wh M)_j^T\,\wh \Theta^{1/2}\wh M\,\wh\Theta_{j}^M =1.
\end{align*}
By the definition of $\wh\Theta_{j}^M$ we also obtain
\begin{align*}
\big\|(\wh \Theta^{1/2}\wh M)_{-j} ^T\wh \Theta^{1/2}\wh M\,\wh\Theta_{j}^M\big\|_\infty
&=\big\|(\wh \Theta^{1/2}\wh M)_{-j} ^T\big((\wh \Theta^{1/2}\wh M)_j - (\wh \Theta^{1/2}\wh M)_{-j}\wh\xi_j\big)\big\|_\infty/\widetilde\tau_j^2\\
&\leq \lambda_j^M/\widetilde\tau_j^2,
\end{align*}
where the last inequality again follows by the KKT conditions for the nodewise Lasso \eqref{eq:gamma:M:Lasso}.
\end{proof}

\begin{proof}[\textsc{ Proof of Proposition \ref{prop:main}.}]
The proof of the first result of the proposition is given in \cite{vandegeer2014}, and hence the proof is omitted. We now prove the second result. The proof relies on the relation
   \begin{align*}
   \|\wh \Theta^M- \Theta^M\|_{op,\infty}&= \max_j \|\wh \Theta^M_j- \Theta^M_j\|_1\\
   &=\max_j \|\wtl \Gamma_j/\widetilde\tau_j^2-\Gamma_j/\tau_j^2\|_1\\
   &\leq \max_j \|\wtl \gamma_j-\gamma_j\|_1/\widetilde\tau_j^2+\max_j\|\gamma_j\|_1 \max_j\left|1/\widetilde\tau_j^2-1/\tau_j^2\right|\\
   &\leq C \Big(\max_j \|\wtl \gamma_j-\gamma_j\|_1\omega  +\omega ^2\sqrt{s_{\max}^M}\max_j\left|\tau_j^2 - \widetilde\tau_j^2\right|\Big) \max_j \frac{1}{\omega \widetilde\tau_j^2},
   \end{align*}
   for all $n$ sufficiently large.
   Here, we made use of the lower bound \eqref{lower:bound:tau} and  $\|\gamma_j\|_1\leq C \sqrt{s_j^M}$  for $n$ sufficiently large.
   We introduce the sets
     \begin{align*}
   \mathcal C_j = \set{  \|\gamma_{S_j}\|_1^2\leq C s_j^M\gamma^T \wh M^T\wh \Theta \wh M\gamma\text{ for all }\|\gamma_{S_j^c}\|_1\leq 3\|\gamma_{S_j}\|_1}
   \end{align*}
   and
   \begin{align*}
   \cT_j=\set{\|\big((\wh\Theta^{1/2}\wh M)_j - (\wh\Theta^{1/2}\wh M)_{-j}\gamma_j\big)^T(\wh \Theta^{1/2}\wh M)_{-j}\|_\infty\leq C\lambda_j^M}
    \end{align*}
    for some sufficiently large constant $C>0$.
    Recall $\lambda_j^M\sim \log(q)/\sqrt{n}$.
   On the set $\mathcal C_j\cap\cT_j$, it holds
      \begin{align*}
   \|\wtl\gamma_j-\gamma_j\|_1\leq C(j) s_j^M \log(q)/\sqrt{n},
   \end{align*}
   for some constant $C(j)>0$,
   which follows directly from Theorem 2.2 of \cite{van2016}.
   Thus, for the proof of the assertion it is sufficient to show
   \begin{align*}
   \left|\widetilde\tau_j^2 - \tau_j^2\right|=O_p\Big(\log(q)\sqrt{s_j^M/n}\Big)
   \end{align*}
   which can be seen as follows.
   Recall from \eqref{Proof:Theorem:4:1:eq:1} that
   \begin{align*}
\wtl{\tau}_j^{2}&=\Big((\wh \Theta^{1/2}\wh M)_j - (\wh \Theta^{1/2}\wh M)_{-j}\wtl\gamma_j\Big)^T
(\wh \Theta^{1/2}\wh M)_j\\
&=\Big((\wh \Theta^{1/2}\wh M)_j - (\wh \Theta^{1/2}\wh M)_{-j}\gamma_j\Big)^T
(\wh \Theta^{1/2}\wh M)_j+\Big((\wh \Theta^{1/2}\wh M)_{-j}(\gamma_j-\wtl\gamma_j)\Big)^T
(\wh \Theta^{1/2}\wh M)_j\\
&=\Big\|(\wh \Theta^{1/2}\wh M)_j - (\wh \Theta^{1/2}\wh M)_{-j}\gamma_j\Big\|_2^2 +\Big((\wh \Theta^{1/2}\wh M)_j - (\wh \Theta^{1/2}\wh M)_{-j}\gamma_j\Big)^T (\wh \Theta^{1/2}\wh M)_{-j}\gamma_j\\
&\quad+\Big((\wh \Theta^{1/2}\wh M)_{-j}(\gamma_j-\wtl\gamma_j)\Big)^T (\wh \Theta^{1/2}\wh M)_j\\
&=\Gamma_j^T \, \wh M^T \wh\Theta \wh M \,\Gamma_j +\Big((\wh \Theta^{1/2}\wh M)_j - (\wh \Theta^{1/2}\wh M)_{-j}\gamma_j\Big)^T (\wh \Theta^{1/2}\wh M)_{-j}\gamma_j\\
&\quad+\Big((\wh \Theta^{1/2}\wh M)_{-j}(\gamma_j-\wtl\gamma_j)\Big)^T (\wh \Theta^{1/2}\wh M)_j
\end{align*}
and recall that $\tau_j^{2} = \Gamma_j^T  M^T \Theta M \Gamma_j$.
We have
   \begin{align*}
   \big|\wtl \tau_j^2-\tau_j^2\big|&\leq
   \underbrace{\big|\Gamma_j^T \big(\wh M^T\wh \Theta\wh M-M^T \Theta M\big) \Gamma_j\big|}_{I}
   +\underbrace{\big|\big((\wh \Theta^{1/2}\wh M)_j - (\wh \Theta^{1/2}\wh M)_{-j}\gamma_j\big)^T
(\wh \Theta^{1/2}\wh M)_{-j}\gamma_j\big|}_{II}\\
 &  +\underbrace{|(\gamma_j-\wtl\gamma_j)^T (\wh \Theta^{1/2}\wh M)_{-j}^T(\wh \Theta^{1/2}\wh M)_j|}_{III}
   \end{align*}
 where we bound each term on the right hand side as follows.
   Consider $I$.
 We observe
\begin{align*}
I&\leq \underbrace{|\Gamma_j^T (\wh M - M)^T \wh\Theta \wh M\,\Gamma_j|}_{T_1}+\underbrace{|\Gamma_j^T M^T(\wh \Theta - \Theta)\wh M\,\Gamma_j |}_{T_2}
+\underbrace{|\Gamma_j^T M^T\Theta(\wh M - M)\,\Gamma_j|}_{T_3}.
\end{align*}
In the following, we bound each summand on the right hand side separately.
We have
\begin{align*}
T_1&=|(\Theta M\Gamma_j)^T(\wh M - M)\Gamma_j|+o_p\big(\log(q)/\sqrt{n}\big)=O_p\big(\log(q)/\sqrt{n}\big),
\end{align*}
uniformly in $j$ by using Assumption \ref{A:M:tech}, i.e.,  $\EE \max_j|(\Theta M\Gamma_j)^TZX^T\Gamma_j|^2=O(\log(p))$, and following the arguments for the upper bound \eqref{upper:bound:III}. Equivalently, we have $T_3=O_p\big(\log(q)/\sqrt{n}\big)$.
We observe
\begin{align*}
T_2
&\leq |\Gamma_j^T M^T\Theta(\wh\Sigma\wh \Theta - I_q)\wh M\,\Gamma_j|+
|\Gamma_j^T M^T\Theta(\wh\Sigma-\Sigma)\wh \Theta \wh M\,\Gamma_j|\\
&\leq |\Gamma_j^T M^T\Theta(\wh\Sigma\wh \Theta - I_q)M\,\Gamma_j|
+|\Gamma_j^T M^T\Theta(\wh\Sigma-\Sigma)\Theta M\,\Gamma_j|
+o_p\big(\sqrt{\log(q)/n}\big)
\end{align*}
Due to Assumption \ref{A:M:tech} $(ii)$, i.e., $\EE \max_{1\leq j\leq p}\|(\Theta M\Gamma_j)^TZ\|_2^4=O(\log(p)^2)$, it  is sufficient to consider the first summand.
 The KKT condition for the nodewise Lasso estimator  $\wh\xi_{j}$ implies
 $\mathbf{Z}_{-j}^T\mathbf{Z}\wh\Theta_j/n = \wh\tau_j^{-2}\lambda_j^{\Theta}\wh\kappa$
and  it holds $\mathbf Z_j^T \mathbf{Z}\wh\Theta_j/n = e_j$ (see \cite{vandegeer2014}).
Consequently, we have
\begin{align*}
\wh\Sigma\wh \Theta_j - e_j =  \lambda_j^{\Theta}\wh\kappa_j /\wh\tau_j^2.
\end{align*}
Since $\lambda_j^{\Theta}\sim\sqrt{\log(q)/n}$ and $\|\Theta M\Gamma_j\|_1\leq \|\Theta\|_1 \|M\|_1\|\Gamma_j\|_1\leq \sqrt{s_{\max}s_M(s_j^M+(\lambda_j^M)^2)}$ we obtain
\begin{align*}
|\Gamma_j^T M^T\Theta(\wh\Sigma\wh \Theta - I_q)M\,\Gamma_j|
&=|\Gamma_j^T M^T\Theta(\lambda_1^{\Theta}\wh\kappa_1 /\wh\tau_1^2,\dots,\lambda_q^{\Theta}\wh\kappa_q /\wh\tau_q^2)M\,\Gamma_j|\\
&\leq\sqrt{\log(q)/n}\,\|\Theta M\Gamma_j\|_1 \|M\,\Gamma_j\|_1\max_{1\leq j\leq q}\wh\tau_j^{-2}\\
&=\sqrt{\log(q)/n}\,\sqrt{s_{\max}}s_M(s_j^M+(\lambda_j^M)^2)\times O_p(1)\\
&=O_p\big(\log(q)/\sqrt{n}\big),
\end{align*}
using that $\wh\tau_j^2$ is a consistent estimator of $1/\Theta_{jj}$ (see the proof of Theorem 2.4 of \cite{vandegeer2014}),  $\Theta_{jj}$ is bounded uniformly in $j$, and the first rate condition imposed in Assumption \ref{A:M} $(iii)$.
 Further, we have on $\cT_j$ that
   \begin{align*}
   II&\leq \|\gamma_j\|_1\big\|\big((\wh\Theta^{1/2}\wh M)_j - (\wh\Theta^{1/2}\wh M)_{-j}\gamma_j\big)^T
(\wh \Theta^{1/2}\wh M)_{-j}\big\|_\infty\\
&=O_p\Big(\log (q)\sqrt{s_j^M/n}\Big).
   \end{align*}
  Further, the  KKT condition for the nodewise Lasso estimator  $\wtl\gamma_{j}$ implies
\begin{align*}
III&=|\Gamma_j^T (\wh \Theta^{1/2}\wh M)^T(\wh \Theta^{1/2}\wh M)_j|\\
&=|\Gamma_j^T (\wh M\wh \Theta\wh M - M\Theta M)e_j|\\
&=O_p\big(\log(q)/\sqrt{n}\big).
\end{align*}   	

  In the following, we show that $\1_{\mathcal C_j\cap \cT_j}$ with probability approaching one.  To control $\mathcal C_j$ we can proceed similarly as in the proof of Theorem \ref{theo:main}. To control $\cT_j$, recall that due to the definition of $\Gamma_j$ it holds  $\Gamma_j^T (\Theta^{1/2}M)^T (\Theta^{1/2}M)_{-j} =0$.
  We observe
\begin{align*}
\|\big((&\wh\Theta^{1/2}\wh M)_j - (\wh\Theta^{1/2}\wh M)_{-j}\gamma_j\big)^T(\wh \Theta^{1/2}\wh M)_{-j}\|_\infty
=\| \Gamma_j^T\big(\wh M^T \wh\Theta \wh M - M^T \Theta M\big) I_{-j} \|_\infty\\
&\leq \underbrace{\|\Gamma_j^T (\wh M - M)^T \wh\Theta \wh M\|_\infty}_{S_1}
+\underbrace{\|\Gamma_j^T M^T(\wh \Theta - \Theta)\wh M\|_\infty }_{S_2}
+\underbrace{\|\Gamma_j^T M^T\Theta(\wh M - M)\|_\infty}_{S_3}.
\end{align*}
In the following, we bound each summand on the right hand side separately.
We have
\begin{align*}
S_1&=\max_l|(\Theta M)_l^T(\wh M - M)\Gamma_j|+o_p\big(\sqrt{\log(q)/n}\big)=O_p\big(\log(q)/\sqrt{n}\big)
\end{align*}
by using Assumption \ref{A:M:tech} $(i)$, i.e., $\EE\|M^T\Theta ZX^T\Gamma_j\|_\infty^2=O(\log(p))$ and following the arguments for the upper bound \eqref{upper:bound:III}.
We observe
\begin{align*}
S_2
%&=\|\Gamma_j^T M^T\Theta(\Sigma\wh \Theta - I_q)M\|_\infty\\
&\leq \|\Gamma_j^T M^T\Theta(\wh\Sigma\wh \Theta - I_q)M\|_\infty+\|\Gamma_j^T M^T\Theta(\wh\Sigma-\Sigma)\wh \Theta M\|_\infty\\
&\leq \|\Gamma_j^T M^T\Theta(\wh\Sigma\wh \Theta - I_q)M\|_\infty+\|\Gamma_j^T M^T\Theta(\wh\Sigma-\Sigma)\Theta M\|_\infty+o_p\big(\sqrt{\log(q)/n}\big)
\end{align*}
where the second summand can be bounded again by using Nemirovski's inequality:
\begin{align*}
\EE\|\Gamma_j^T M^T\Theta(\wh\Sigma-\Sigma)\Theta M\|_\infty^2&
=\EE\max_{1\leq l\leq p} \big|n^{-1}\sum_i \Gamma_j^T M^T\Theta Z_i Z_i^T(\Theta M)_l - \Gamma_j^T M^T(\Theta M)_l\big|^2\\
&\leq 8\log(2p)n^{-1}  \EE \max_{1\leq l\leq p}|\Gamma_j^T M^T\Theta Z Z^T(\Theta M)_l|^2.
\end{align*}
 The KKT condition for the nodewise Lasso estimator  $\wh\xi_{j}$ implies
 $\mathbf{Z}_{-j}^T\mathbf{Z}\wh\Theta_j/n = \wh\tau_j^{-2}\lambda_j^{\Theta}\wh\kappa_j$
and  it holds $\mathbf Z_j^T \mathbf{Z}\wh\Theta_j/n = e_j$.
Consequently, we have
\begin{align*}
\wh\Sigma\wh \Theta_j - e_j =  \lambda_j^{\Theta}\wh\kappa_j /\wh\tau_j^2.
\end{align*}
 Since $\lambda_j^{\Theta}\sim\sqrt{\log(q)/n}$ we obtain by employing Theorem 2.4 of \cite{vandegeer2014})
\begin{align*}
\|\Gamma_j^T M^T\Theta(\wh\Sigma\wh \Theta - I_q)M\|_\infty&=\|\Gamma_j^T M^T\Theta(\lambda_1^{\Theta}\wh\kappa_1 /\wh\tau_1^2,\dots,\lambda_q^{\Theta}\wh\kappa_q /\wh\tau_q^2)M\|_\infty\\
%&= \sqrt{\log(q)/n}\, \|\Gamma_j^T M^T\Theta(\widehat \kappa_1,\dots,\widehat \kappa_q)\text{diag}(\Theta)M\|_\infty+o_p(1)\\
&\leq\sqrt{\log(q)/n}\,\|\Theta M\Gamma_j\|_1 \|M\|_1\max_{1\leq j\leq q}\wh\tau_j^{-2}\\
&=\sqrt{\log(q)/n}\,\sqrt{s_{\max}}s_M\sqrt{s_j^M+(\lambda_j^M)^2}\times O_p(1)\\
&=O_p(\log(q)/\sqrt{n}),
\end{align*}
by using Assumption \ref{A:M} $(iii)$, i.e., $s_M\sqrt{s_{\max} s_{\max}^M}=O\big(\sqrt{\log(q)}\big)$.
Finally, we have
\begin{align*}
S_3&=\|(\Theta M\Gamma_j)^T(\wh M - M)\|_\infty=O_p\big(\log(q)/\sqrt{n}\big),
\end{align*}
by following again the arguments for the upper bound \eqref{upper:bound:III}, which completes the proof of the result.
\end{proof}

%========================================================================
%========================================================================
For a random variable $W$, we introduce  the sub-Gaussian norm $\|\cdot\|_{\psi_2}$ as $\|W\|_{\psi_2}:=\sup_{q\geq 1} q^{-1/2}(\EE|W|^q)^{1/q}$ and the  sub-exponential norm $\|\cdot \|_{\psi_1}$ as $\|W\|_{\psi_1}:=\sup_{q\geq 1} q^{-1}(\EE|W|^q)^{1/q}$, see \cite[Definition 5.7 and Lemma 5.5]{Vershynin2012}. If $W$ is sub-Gaussian (see Definition \ref{A:Z}) then $\|W\|_{\psi_2}$ is bounded from above. Also note that if $W$ has bounded sub-Gaussian norm then $W^2$ has bounded
 sub-exponential norm, see \cite[Remark 5.18]{Vershynin2012}.
\begin{proof}[\textsc{ Proof of Proposition \ref{th:norm1:M}.}]
  We start by proving the first part of the theorem. %We only prove the first part of the theorem as the second part follows from the proof of \cite[Theorem 4]{CaiZhou2012}.
  Denote $\varsigma_{zj}^2:=\EE[\mb Z_{1j}^2]$, $\varsigma_{xk}^2:=\EE[\mb X_{1k}^2]$ and $\rho_{jk}:=\EE[\mb Z_{1j} \mb X_{1k}]/(\varsigma_{zj}\varsigma_{xk})$.  Let $K_z:= \|\mb Z_{ij}\|_{\psi_2}$ and $K_x:= \|\mb X_{ik}\|_{\psi_2}$, which do not depend on $i$. Then,
\begin{eqnarray*}
  \P\left(\left|\wtl M_{jk} - M_{jk}\right| \geq v \right) & = & \P\left(\left|\sum_{i=1}^n \left(\mb Z_{ij} \mb X_{ik} - M_{jk}\right)\right| \geq nv \right)\\
  & = &
   \P\left(\left|\sum_{i=1}^n \left(\frac{\mb Z_{ij} \mb X_{ik}}{\varsigma_{zj}\varsigma_{xk}} - \rho_{jk}\right)\right| \geq \frac{nv}{\varsigma_{zj}\varsigma_{xk}} \right).
\end{eqnarray*}
Moreover, %as in \cite[Lemma A.3]{BickelLevina2008a}
\begin{multline*}
  \sum_{i=1}^n \left(\frac{\mb Z_{ij} \mb X_{ik}}{\varsigma_{zj}\varsigma_{xk}} - \rho_{jk}\right) = \frac{1}{4}\Big[\sum_{i=1}^n [\left(\frac{\mb Z_{ij}}{\varsigma_{zj}} + \frac{\mb X_{ik}}{\varsigma_{xk}}\right)^2 - 2(1 + \rho_{jk})]\\
   - \sum_{i=1}^n [\left(\frac{\mb Z_{ij}}{\varsigma_{zj}} - \frac{\mb X_{ik}}{\varsigma_{xk}}\right)^2 - 2(1 - \rho_{jk})]\Big].
\end{multline*}
Because $\mb X$ and $\mb Z$ have sub-Gaussian rows then $\mb X_{ik}$, $\mb Z_{ij}$, $\left(\frac{\mb Z_{ij}}{\varsigma_{zj}} + \frac{\mb X_{ik}}{\varsigma_{xk}}\right)$ and $\left(\frac{\mb Z_{ij}}{\varsigma_{zj}} - \frac{\mb X_{ik}}{\varsigma_{xk}}\right)$ are sub-Gaussian (because linear combinations of sub-Gaussian random variables are still sub-Gaussian). The sub-gaussian norms of $\left(\frac{\mb Z_{ij}}{\varsigma_{zj}} + \frac{\mb X_{ik}}{\varsigma_{xk}}\right)$ and $\left(\frac{\mb Z_{ij}}{\varsigma_{zj}} - \frac{\mb X_{ik}}{\varsigma_{xk}}\right)$ are upper bounded by $\frac{K_z}{\zeta_{zj}} + \frac{K_x}{\zeta_{xk}}$.

Therefore, $\left(\frac{\mb Z_{ij}}{\zeta_{zj}} + \frac{\mb X_{ik}}{\zeta_{xk}}\right)^2$ and $\left(\frac{\mb Z_{ij}}{\zeta_{zj}} - \frac{\mb X_{ik}}{\zeta_{xk}}\right)^2$ are sub-exponential, see \textit{e.g.} \cite[Lemma 5.14]{Vershynin2012}, whose means are, respectively
$$2(1 + \rho_{jk})\qquad \textrm{ and } \qquad 2(1 - \rho_{jk}).$$
Denote $W_{i+}:= \left(\frac{\mb Z_{ij}}{\zeta_{zj}} + \frac{\mb X_{ik}}{\zeta_{xk}}\right)^2\frac{1}{2(1 + \rho_{jk})} - 1$ and $W_{i-}:= \left(\frac{\mb Z_{ij}}{\zeta_{zj}} - \frac{\mb X_{ik}}{\zeta_{xk}}\right)^2\frac{1}{2(1 -\rho_{jk})} - 1$ which are also sub-exponential by \cite[Remark 5.18]{Vershynin2012} with mean zero. In fact, by using the moment condition characterization of sub-Gaussianity we obtain, for some constant $K>0$ and all $p\geq 1$:
\begin{multline*}
  (\EE|W_{i+}|^p)^{1/p} \leq \left\|\left(\frac{\mb Z_{ij}}{\zeta_{zj}} + \frac{\mb X_{ik}}{\zeta_{xk}}\right)^2\frac{1}{2(1 + \rho_{jk})}\right\|_p + \|1\|_p\\
  \leq 2\left\|\left(\frac{\mb Z_{ij}}{\zeta_{zj}} + \frac{\mb X_{ik}}{\zeta_{xk}}\right)^2\frac{1}{2(1 + \rho_{jk})}\right\|_p \leq 2K p
\end{multline*}
where we have use the triangle inequality to get the first inequality, the Jensen inequality to get the second inequality and sub-exponentiality of $\left(\frac{\mb Z_{ij}}{\zeta_{zj}} + \frac{\mb X_{ik}}{\zeta_{xk}}\right)^2\frac{1}{2(1 + \rho_{jk})}$ to get the last inequality.\\
The sub-exponential norm of $W_{i+}$ can be upper bounded as follows:
\begin{multline}
  \|W_{i+}\|_{\psi_1} \leq \sup_{q \geq 1} q^{-1} \|W_{i+}\|_q \leq \sup_{q \geq 1} q^{-1} \left(\left\|\left(\frac{\mb Z_{ij}}{\zeta_{zj}} + \frac{\mb X_{ik}}{\zeta_{xk}}\right)^2\frac{1}{2(1 + \rho_{jk})}\right\|_q + \|1\|_q\right)\\
  \leq \left\|\left(\frac{\mb Z_{ij}}{\zeta_{zj}} + \frac{\mb X_{ik}}{\zeta_{xk}}\right)^2\frac{1}{2(1 + \rho_{jk})}\right\|_{\psi_1} + \sup_{q \geq 1} q^{-1}\left\|\left(\frac{\mb Z_{ij}}{\zeta_{zj}} + \frac{\mb X_{ik}}{\zeta_{xk}}\right)^2\frac{1}{2(1 + \rho_{jk})}\right\|_q\\\hfill
  = 2 \left\|\left(\frac{\mb Z_{ij}}{\zeta_{zj}} + \frac{\mb X_{ik}}{\zeta_{xk}}\right)^2\frac{1}{2(1 + \rho_{jk})}\right\|_{\psi_1} \leq 4 \left\|\left(\frac{\mb Z_{ij}}{\zeta_{zj}} + \frac{\mb X_{ik}}{\zeta_{xk}}\right)\frac{1}{\sqrt{2(1 + \rho_{jk})}}\right\|_{\psi_2}^2\\
  \leq 4 \left(\frac{K_z}{\zeta_{zj}} + \frac{K_x}{\zeta_{xk}}\right)^2 \frac{1}{2(1 + \rho_{jk})}
\end{multline}
\noindent where we have first used the triangle inequality, then the Jensen's inequality and, to get the third inequality we have used \cite[Lemma 5.14]{Vershynin2012}. In a similar way, we can show that the sub-exponential norm of $W_{i-}$ is upper bounded by
\begin{equation}
  \|W_{i-}\|_{\psi_1} \leq 4 \left(\frac{K_z}{\zeta_{zj}} + \frac{K_x}{\zeta_{xk}}\right)^2 \frac{1}{2(1 - \rho_{jk})}
\end{equation}
\noindent and the right hand side does not depend on $i$. Therefore, for every $i$, $\|(1 + \rho_{jk}) W_{i+}\|_{\psi_1}\leq 2\left(\frac{K_z}{\zeta_{zj}} + \frac{K_x}{\zeta_{xk}}\right)^2$ and $\|(1 - \rho_{jk}) W_{i-}\|_{\psi_1} \leq 2\left(\frac{K_z}{\zeta_{zj}} + \frac{K_x}{\zeta_{xk}}\right)^2$. Let $K:=\max_i\|(1 - \rho_{jk}) W_{i-}\|_{\psi_1}$. For every $t \geq 0$, define the event $\mathcal{A}:=\{|\sum_{i=1}^n (1 - \rho_{jk}) W_{i-}| \geq t\}$ which by using \cite[Proposition 5.16]{Vershynin2012} has probability upper bounded by
\begin{multline}\label{proof:th_4.4:eq1}
  \P(\mathcal{A}) \leq 2\exp\left\{-c\min\left\{\frac{t^2}{K^2 n},\frac{t}{K}\right\}\right\}\\
  \leq 2\exp\left\{-c\min\left\{\frac{t^2}{4n\left(\frac{K_z}{\zeta_{zj}} + \frac{K_x}{\zeta_{xk}}\right)^4},\frac{t}{2\left(\frac{K_z}{\zeta_{zj}} + \frac{K_x}{\zeta_{xk}}\right)^2}\right\}\right\}
\end{multline}
where $c>0$ is an absolute constant. The probability that we want to upper bound is the following:
\begin{multline*}
  \P\left(\left|\wtl{M}_{jk} - M_{jk}\right| \geq v \right) = \P\left(\frac{1}{2}\left|\sum_{i=1}^n W_{i+} (1 + \rho_{jk}) - \sum_{i=1}^n W_{i-} (1 - \rho_{jk})\right| \geq \frac{nv}{\zeta_{zj}\zeta_{xk}} \right)\\
  \leq \P\left(\left|\sum_{i=1}^n W_{i+} (1 + \rho_{jk})\right| \geq 2\frac{nv}{\zeta_{zj}\zeta_{xk}} - \left|\sum_{i=1}^n W_{i-} (1 - \rho_{jk})\right| \cap \mathcal{A}^c\right) + \P(\mathcal{A})\\
  \leq \P\left(\left|\sum_{i=1}^n W_{i+} (1 + \rho_{jk})\right| \geq \frac{nv}{\zeta_{zj}\zeta_{xk}}\cap \mathcal{A}^c \right) + \P(\mathcal{A})\\
  \leq \P\left(\left|\sum_{i=1}^n W_{i+} (1 + \rho_{jk})\right| \geq \frac{nv}{\zeta_{zj}\zeta_{xk}}\right) + \P(\mathcal{A}).
\end{multline*}
Therefore, by using \eqref{proof:th_4.4:eq1} with $t = nv/(\zeta_{zj}\zeta_{xk})$ and $0\leq v \leq 1$ in $\mathcal{A}$, and applying again \cite[Proposition 5.16]{Vershynin2012} to upper bound the first probability in the last line of the previous display, we obtain
\begin{multline*}
  \P\left(\left|\wtl{M}_{jk} - M_{jk}\right| \geq v \right)\\
   \leq 2\exp\left\{-c\min\left\{\frac{v^2}{4\zeta_{zj}^2\zeta_{xk}^2\left(\frac{K_z}{\zeta_{zj}} + \frac{K_x}{\zeta_{xk}}\right)^4},\frac{v}{2\zeta_{zj}\zeta_{xk}\left(\frac{K_z}{\zeta_{zj}} + \frac{K_x}{\zeta_{xk}}\right)^2}\right\}n\right\}\\\hfill
  + 2\exp\left\{-c\min\left\{\frac{v^2}{4\zeta_{zj}^2\zeta_{xk}^2\left(\frac{K_z}{\zeta_{zj}} + \frac{K_x}{\zeta_{xk}}\right)^4},\frac{v}{2\zeta_{zj}\zeta_{xk}\left(\frac{K_z}{\zeta_{zj}} + \frac{K_x}{\zeta_{xk}}\right)^2}\right\}n\right\}\\
  = 4\exp\left\{-c\min\left\{\frac{v^2}{4\zeta_{zj}^2\zeta_{xk}^2\left(\frac{K_z}{\zeta_{zj}} + \frac{K_x}{\zeta_{xk}}\right)^4},\frac{v}{2\zeta_{zj}\zeta_{xk}\left(\frac{K_z}{\zeta_{zj}} + \frac{K_x}{\zeta_{xk}}\right)^2}\right\}n\right\}\\
  \leq 4 \exp\{-C v^2n\}
\end{multline*}
where for the last inequality we have used that$\min(a/b,c/d) \geq \min(a,c)/\max(b,d)$ for any constants $a,b,c,d$. %$C_5 := c\zeta_{zj}\zeta_{xk}[2(K_z\zeta_{xk} + K_x\zeta_{zj})^2 \max(2(K_z\zeta_{xk} + K_x\zeta_{zj})^2/(\zeta_{zj}\zeta_{xk}),1)]^{-1}$.
This proves \eqref{lem:result:1}.
To prove the second part of the theorem notice that, by definition of $s_M$ and under Assumption \ref{A:M} \textit{(iii)}, $M$ belongs to the class of matrices
\begin{equation}\label{eq_class_covariance}
  \mathcal{G}_\chi(\rho,s_M) = \left\{M \in\mathbb{R}^{q\times p}: \, \max_{1 \leq k\leq p}|M_{[j]k}|^\chi \leq s_M/j \text{ for all } j \text{ and }\max_{1 \leq j\leq (p\wedge q)}M_{jj} \leq \rho\right\}
\end{equation}
with $\chi = 0$ and $|M_{[j]k}|$ denoting the $j$-th largest element in magnitude of the $k$-th column $(M_{jk})_{1\leq j \leq q}$ of $M$. This is the extension to rectangular matrices of the class of matrices considered in \cite{CaiZhou2012} for $0\leq \chi < 1$. Hence, the second part of the theorem follows from the proof of \cite[Theorem 4]{CaiZhou2012} and \eqref{lem:result:1}. We give some elements of this proof in Appendix \ref{A:technical:results}.
\end{proof}

\section{Appendix: Technical Results}\label{A:technical:results}
Recall the notation $\wtl M = \mathbf Z^T\mathbf X/n$ and the thresholding estimator: $\wh{M} = (\wh{M}_{jk})$ with
\begin{equation}
  \wh{M}_{jk} := \wtl M_{jk}\1\left\{|\wtl M_{jk}| \geq C_0 \sqrt{\frac{\log (q)}{n}}\right\},\qquad C_0 > 0.
\end{equation}
\noindent In the following theorem, we provide the rate for its $\ell_1$-norm. The minimax rate for the $\ell_1$-norm of the thresholding estimator of quadratic matrix is studied in \cite{CaiZhou2012}. Here, we slightly extend their proof to account for the rectangular case and only report the main steps that contain the differences with respect to \cite{CaiZhou2012}. We will establish this result for the more general class of matrices  $\mathcal{G}_\chi(\rho,s_M)$ defined in \eqref{eq_class_covariance} for $0 \leq \chi <1$
 where $|M_{[j]k}|$ denotes the $j$-th largest element in magnitude of the $k$-th column $(M_{jk})_{1\leq j \leq q}$. Every matrix in $\mathcal{G}_\chi(\rho,s_M)$ has columns $(M_{jk})_{1\leq j \leq q}$ that are in a (approximate) sparse weak $\ell_\chi$ ball. The case $\chi = 0$ is the case considered in the paper. Moreover, define the class of distributions $\mathcal{P}(\mathcal{G}_\chi(\rho,s_M))$ as the set of distributions of $(Z,X)$ satisfying \eqref{eq_class_covariance} and such that the rows of $\mathbf Z$ and $\mathbf X$ are sub-Gaussian.
\begin{theo}\label{tr:technical:appendix}
  Let Assumption \ref{A:Z} $(ii)$ hold. Then, the thresholding estimator $\wh M$ satisfies
  \begin{displaymath}
    \sup_{\mathcal{P}(\mathcal{G}_\chi(\rho,s_M))}\EE \big\|\wh M - M\big\|_1^2 \leq C s_M^2 \left(\frac{\log(p\vee q)}{n}\right)^{1 - \chi}
  \end{displaymath}
  for some constant $C>0$.
\end{theo}
In the following we directly write $q$ instead of $p\vee q$. Therefore, by Theorem \ref{tr:technical:appendix} and the Markov's inequality
\begin{eqnarray}
  \P\left(\left\|\wh M - M\right\|_1 > \varepsilon\right) & \leq & \frac{1}{\varepsilon^2}\EE \left\|\wh M - M\right\|_1^2\nonumber\\
  & \leq & C\frac{s_M^2}{\varepsilon^2}\left(\frac{\log(q)}{n}\right)^{1 - \chi}
\end{eqnarray}
which implies:
$$\left\|\wh M - M\right\|_1 \leq s_M \left(\frac{\log(q)}{n}\right)^{(1 - \chi)/2}$$
\noindent with probability approaching one.
\proof
Define the event $A_{jk}:=\{|\wh{M}_{jk} - M_{jk}| \leq 4 \min\left\{|M_{jk}|,C_0\sqrt{\frac{\log(q)}{n}}\right\}\}$ and $D = (d_{jk})$ with $d_{jk} := (\wh{M}_{jk} - M_{jk}) \1_{A_{jk}^c}$. Then,
\begin{eqnarray}
  \EE \|\wh M - M\|_1^2 & = & \EE \|\wh M - M - D + D\|_1^2\nonumber\\
  & \leq & \EE \|\wh M - M - D\|_1^2 + \EE\|D\|_1^2\nonumber\\
  & \leq & 2 \EE\left(\sup_{1 \leq k \leq p} \sum_{j=1}^q |\wh M_{jk} - M_{jk}|\1_{A_{jk}}\right)^2 + 2 \EE\|D\|_1^2\nonumber\\
  & \leq & 32 \left(\sup_{1\leq k\leq p}\sum_{j=1}^q \min \left\{|M_{jk}|,C_0 \sqrt{\frac{\log(q)}{n}}\right\}\right)^2 + 2 \EE\|D\|_1^2\label{proof:1:eq:1}
\end{eqnarray}
where the inequality in the penultimate line is due to $(\wh M - M - D)_{jk} = (\wh{M}_{jk} - M_{jk})(1 - \1_{A_{jk}^c}) = (\wh{M}_{jk} - M_{jk})\1_{A_{jk}}$.\\
To control the first term we use exactly the same procedure as in \cite{CaiZhou2012} and so we omit it. We find that
\begin{equation}
  32 \left(\sup_{1\leq k\leq p}\sum_{j=1}^q \min \left\{|M_{jk}|,C_0 \sqrt{\frac{\log(q)}{n}}\right\}\right)^2 \leq C_1 s_M \left(\frac{\log(q)}{n}\right)^{(1 - \chi)/2}
\end{equation}
\noindent for some positive constant $C_1$. We now consider the second term in \eqref{proof:1:eq:1} and show that it is negligible with respect to the first term.
For this we use the following decomposition (also coming from \cite{CaiZhou2012}), where we denote by $\|\cdot\|_F$ the Frobenius norm:
\begin{eqnarray*}
  \EE\|D\|_1^2 & = & \EE \left(\max_{1 \leq k \leq p}\sum_{j=1}^q|d_{jk}|\right)^2 \leq \EE[q \|D\|_F^2] = q\sum_{k=1}^p\sum_{j=1}^q \EE|d_{jk}|^2\\
  & = & q\sum_{k=1}^p\sum_{j=1}^q \EE\left(d_{jk}^2 \1_{\{A_{jk}^c\cap\{\wh M_{jk} = \wtl M_{jk}\}\}} + d_{jk}^2 \1_{\{A_{jk}^c\cap\{\wh M_{jk} = 0\}\}}\right)\\
  & = & q\sum_{k=1}^p\sum_{j=1}^q \EE\left((\wtl M_{jk} - M_{jk})^2 \1_{\{A_{jk}^c\}} + M_{jk}^2 \1_{\{A_{jk}^c\cap\{\wh M_{jk} = 0\}\}}\right) =: R_1 + R_2.
\end{eqnarray*}
Let us start by term $R_1$. By the Holder's inequality (with norms $L_3$ and $L_{3/2}$) we obtain
\begin{eqnarray*}
  R_1 & \leq & p\sum_{k=1}^p\sum_{j=1}^q \EE^{1/3}\left[(\wtl M_{jk} - M_{jk})^6\right] \P^{2/3}(A_{jk}^c)\\
  & \leq & C_3 p^2 q\frac{1}{n}\P^{2/3}(A_{jk}^c)
\end{eqnarray*}
where we have used result \eqref{lem:result:2} of Lemma \ref{lem:1} below that $\EE^{1/3}\left[(\wtl M_{jk} - M_{jk})^6\right] = O (n^{-1})$. Finally, by using the result of Lemma \ref{lem:2} below we get that $\P(A_{jk}^c) \leq 2 C_4q^{-9/2}$ so that
$$R_1 \leq 2 \frac{C_2 C_3}{n} q^3 q^{-3} \leq C_5/n.$$
\indent Let us now consider term $R_2$:
\begin{eqnarray*}
  R_2 & = & p\sum_{k=1}^p\sum_{j=1}^q\EE\left(M_{jk}^2\1_{\{|M_{jk}| \geq 4 C_0\sqrt{\log (q)/n}\}}\1_{\{|\wtl M_{jk}| \leq C_0\sqrt{\log (q)/n}\}}\right)\\
  & \leq & p\sum_{k=1}^p\sum_{j=1}^qM_{jk}^2\EE\left(\1_{\{|M_{jk}| \geq 4 C_0\sqrt{\log (q)/n}\}}\1_{\{|M_{jk}| - |\wtl M_{jk} - M_{jk}| \leq C_0\sqrt{\log (q)/n}\}}\right)\\
  & = & \frac{p}{n}\sum_{k=1}^p\sum_{j=1}^q n M_{jk}^2\P\left(|\wtl M_{jk} - M_{jk}| \geq - C_0\sqrt{\log (q)/n} + |M_{jk}|\right)\1_{\{|M_{jk}| \geq 4 C_0\sqrt{\log (q)/n}\}}\\
  & \leq & \frac{p}{n}\sum_{k=1}^p\sum_{j=1}^q n M_{jk}^2\P\left(|\wtl M_{jk} - M_{jk}| \geq - \frac{1}{4}|M_{jk}| + |M_{jk}|\right)\1_{\{|M_{jk}| \geq 4 C_0\sqrt{\log (q)/n}\}}
\end{eqnarray*}
\noindent where to get the inequality in the second line we have used $|\wtl M_{jk}| \geq |M_{jk}| - |\wtl M_{jk} - M_{jk}|$. Therefore, by using result \eqref{lem:result:1} in Theorem \ref{th:norm1:M} we get:
\begin{eqnarray*}
  R_2 & \leq & \frac{p}{n}\sum_{k=1}^p\sum_{j=1}^q n M_{jk}^2\P\left(|\wtl M_{jk} - M_{jk}| \geq \frac{3}{4}|M_{jk}|\right)\1_{\{|M_{jk}| \geq 4 C_0\sqrt{\log (q)/n}\}}\\
  & \leq & \frac{pC_0^2}{n}\sum_{k=1}^p\sum_{j=1}^q \frac{n}{C_0^2} M_{jk}^2 4 \exp\{-c n 9|M_{jk}|^2/16\}\1_{\{|M_{jk}| \geq 4 C_0\sqrt{\log (q)/n}\}}\\
  & \leq & \frac{pC_0^2}{n}\sum_{k=1}^p\sum_{j=1}^q \exp\{n M_{jk}^2 \frac{4}{C_0^2} - c n 9|M_{jk}|^2/16\}\1_{\{|M_{jk}| \geq 4 C_0\sqrt{\log (q)/n}\}}
\end{eqnarray*}
where to get the last inequality we have used the inequality $nt/e^{nt} \leq 1$ for all $t>0$. Let $C_0 = \sqrt{8/c}$, then
\begin{eqnarray*}
  R_2 & \leq & \frac{pC_0^2}{n}\sum_{k=1}^p\sum_{j=1}^q \exp\{-n M_{jk}^2 c/16\}\1_{\{|M_{jk}| \geq 4 C_0\sqrt{\log (q)/n}\}}\\
  & \leq & \frac{pC_0^2}{n}\sum_{k=1}^p\sum_{j=1}^q \exp\{-n 16C_0^2\log (q) c/(16n)\}\\
  & \leq & \frac{pC_0^2}{n}\sum_{k=1}^p\sum_{j=1}^q \exp\{-8\log (q) \} = \frac{(q)^3C_0^2}{n} (q)^{-8}\\
  & \leq & C_6/n.
\end{eqnarray*}

\begin{lem}\label{lem:1}
  Let Assumption \ref{A:Z} $(ii)$ hold. Then, there exists a constant $C_2>0$ such that
  \begin{equation}\label{lem:result:2}
    \EE[|\wtl M_{jk}- M_{jk}|^6] \leq \frac{24}{C_2^3 n^3}.
  \end{equation}
\end{lem}
\proof
The $6$-th moment can be written as
\begin{displaymath}
  \EE[|\wtl M_{jk}- M_{jk}|^6] = 6\int_{0}^{\infty} x^5 \P(|\wtl M_{jk}- M_{jk}| \geq x) dx
\end{displaymath}
and by substituting the upper bound in \eqref{lem:result:1} and by using integration by parts we get
\begin{eqnarray*}
  \EE[|\wtl M_{jk}- M_{jk}|^6] & \leq & 24 \int_{0}^{\infty} x^5 \exp\left\{- c nx^2\right\}dx = \frac{24}{c^3 n^3}.
\end{eqnarray*}

\begin{lem}\label{lem:2}
  Define the event $A_{jk}$ as $A_{jk} := \Big\{|\wh{M}_{jk} - M_{jk}| \leq 4 \min\big\{|M_{jk},C_0\sqrt{\frac{\log(q)}{n}}|\big\}\Big\}$
  for $C_0 = \sqrt{\frac{8}{C_2}}$ where $C_2$ is as in Lemma \ref{lem:1}. Then,
  $$\P(A_{jk}) \geq 1 - 2 C_3(q)^{-9/2}$$
  for some constant $C_3 > 0$.
\end{lem}

\proof Let $A_1:= \left\{|\wtl M_{jk}| \geq C_0 \sqrt{\frac{\log (p \vee q)}{n}}\right\}$. Then, from the definition of $\wh{M}_{jk}$ we have
\begin{displaymath}
  |\wh{M}_{jk} - M_{jk}| = |M_{jk}|\1_{A_1^c} + |\wtl M_{jk}- M_{jk}|\1_{A_1}.
\end{displaymath}
By the triangular inequality we have:
\begin{eqnarray*}
  A_1 & = & \left\{|\wtl M_{jk}- M_{jk} + M_{jk}| \geq C_0 \sqrt{\frac{\log(q)}{n}}\right\} \subset \left\{|\wtl M_{jk}- M_{jk}| \geq C_0 \sqrt{\frac{\log(q)}{n}} - | M_{jk}|\right\}\\
  A_1^c & = & \left\{|\wtl M_{jk}- M_{jk} + M_{jk}| < C_0 \sqrt{\frac{\log(q)}{n}}\right\} \subset \left\{|\wtl M_{jk}- M_{jk}| > | M_{jk}| - C_0 \sqrt{\frac{\log(q)}{n}}\right\}.
\end{eqnarray*}
Then, the proof proceed exactly as in \cite[Proof of Lemma 8]{CaiZhou2012} with $C_0 = \sqrt{\frac{8}{C_2}}$ where $C_2$ is as in the statement of Lemma \ref{lem:1}.\\

\section{Methodology used for the cross-validation}\label{A:cross:validation}
Implementation of our procedure requires the choice of tuning parameters, namely $\lambda$, $\lambda_j^{\Theta}$, $j = 1,\ldots, q$, $\lambda_j^M$, $j=1,\ldots, p$, and $C_0$. These parameters have been chosen by 10-fold cross-validation in our numerical implementation of our procedure. In this section we describe the precise methodology that we have used for the cross-validation.\\
\indent Consider first the cross-validation procedure to choose $\lambda$ in the construction of the IV Lasso estimator $\wtl\beta$ in \eqref{eq:gamma:beta:Lasso}. The algorithm is the following.\\

\textsc{Algorithm 1.}
  \begin{itemize}
    \item Randomly divide the set of indices $\{1,\ldots, q\}$ into $10$ groups, or folds, of approximately equal size.
    \item For $i=1,\ldots,10$:
    \begin{enumerate}
      \item construct a submatrix that contains only the rows of $\wh\Theta^{1/2}$ corresponding to the indices in the $i$-th fold and denote it by $(\wh\Theta^{1/2})^{(i)}$;
      \item construct a submatrix that contains all the rows of $\wh\Theta^{1/2}$ except the ones corresponding to the indices in the $i$-th fold and denote it by $(\wh\Theta^{1/2})^{(-i)}$;
      %submatrices, say $(\wh\Theta^{1/2}\mathbf Z^T\mathbf Y/n)^{(i)}$ and $(\wh\Theta^{1/2}\wh M)^{(i)}$, that contain only the rows of $\wh\Theta^{1/2}\mathbf Z^T\mathbf Y/n$ and of $\wh\Theta^{1/2}\wh M$, respectively, corresponding to the indices in the $i$-th fold;
      %\item similarly construct submatrices, say $(\wh\Theta^{1/2}\mathbf Z^T\mathbf Y/n)^{(-i)}$ and $(\wh\Theta^{1/2}\wh M)^{(-i)}$, that do not contain the rows of $\wh\Theta^{1/2}\mathbf Z^T\mathbf Y/n$ and of $\wh\Theta^{1/2}\wh M$, respectively, corresponding to the indices in the $i$-th fold;
      \item for a given $\lambda$, solve the minimization problem in \eqref{eq:gamma:beta:Lasso} with these submatrices:
      $$\wtl\beta^{(-i)}(\lambda) = \argmin_{\beta\in\mathbb{R}^{p}}\left\{\|(\wh\Theta^{1/2})^{(-i)} \,(\mathbf Z^T\mathbf Y/n - \wh M \beta)\|_2^2 +  2\lambda\,\|\beta\|_1\right\}.$$
      This gives $\wtl\beta^{(-i)}(\lambda)$;
      \item compute the mean squared error $MSE^{(-i)}(\lambda)$ associated to the given $\lambda$ as:
        \begin{displaymath}
          MSE^{(-i)}(\lambda) := \|(\wh\Theta^{1/2})^{(i)}(\mathbf Z^T\mathbf Y/n - \wh M\wtl\beta^{(-i)}(\lambda))\|_2^2.
        \end{displaymath}
    \end{enumerate}
    \item Compute the $10$-fold cross-validation estimate for the test mean squared error as
        \begin{equation}
          CV_{10}(\lambda) = \frac{1}{10}\sum_{i=1}^{10} MSE^{(-i)}(\lambda).
        \end{equation}
    \item Choose the $\lambda$ that minimizes $CV_{10}(\lambda)$.
  \end{itemize}
%\textsc{Algorithm 1.}
%  \begin{itemize}
%    \item randomly divide the set of observations $(Y_1,X_1,Z_1),\dots,(Y_n,X_n,Z_n)$ into $10$ groups, or folds, of approximately equal size.
%    \item For $i=1,\ldots,10$:
%    \begin{enumerate}
%      \item use the $i$-th fold as a validation set, and solve the minimization problem in \eqref{eq:gamma:beta:Lasso} for a given $\lambda$ by using only observations in the remaining $9$ folds. This gives $\wtl\beta^{(-i)}(\lambda)$;
%      \item compute $\wh M$ and $\wh\Theta$ by using the observations in the held-out fold $i$ and denote them by $\wh M^{(-i)}$ and $\wh\Theta^{(-i)}$.
%      \item let $\mathbf Z_{(-i)}$ and $\mathbf Y_{(-i)}$ be the matrix and vector, respectively, of observations in the held-out fold $i$ and let $n_i$ be the number of observations in the held-out fold $i$.
%      \item compute the mean squared error $MSE^{(-i)}(\lambda)$ associated to a given $\lambda$ as:
%        \begin{multline}
%          MSE^{(-i)}(\lambda) := \\
%          \left(\frac{\mathbf Z_{(-i)}^T\mathbf Y_{(-i)}}{n_i} - \wh M^{(-i)}\wtl\beta^{(-i)}(\lambda)\right)^T\wh\Theta^{(-i)} \left(\frac{\mathbf Z_{(-i)}^T\mathbf Y_{(-i)}}{n_i} - \wh M^{(-i)} \wtl\beta^{(-i)}(\lambda)\right).
%        \end{multline}
%    \end{enumerate}
%    \item Compute the $10$-fold cross-validation estimate for the test mean square error as
%        \begin{equation}
%          CV_{10}(\lambda) = \frac{1}{10}\sum_{i=1}^{10} MSE^{(-i)}(\lambda).
%        \end{equation}
%    \item Choose the $\lambda$ that minimizes $CV_{10}(\lambda)$.
%  \end{itemize}
In practice one has to use a grid for $\lambda$ and select the value in this grid that gives a minimum value for $CV_{10}(\lambda)$. This procedure is automatically produced by the \texttt{R} function \verb"cv.glmnet" of the \verb"glmnet" package.\\
\indent The cross-validation procedure to choose $\lambda_j^{\Theta}$, $j = 1,\ldots, q$, is described in the following algorithm.\\

\textsc{Algorithm 2.}
  \begin{itemize}
    \item Randomly divide the set of observations $Z_1,\dots,Z_n$ into $10$ groups, or folds, of approximately equal size.
    \item For $i=1,\ldots,10$:
      \begin{enumerate}
        \item construct a subvector and a submatrix of $\mathbf Z_{j}$ and $\mathbf Z_{-j}$ that contain only the observations in the held-out $i$-th fold and denote them by $\mathbf Z_{j}^{(i)}$ and $\mathbf Z_{-j}^{(i)}$, respectively;
        \item construct a subvector and a submatrix of $\mathbf Z_{j}$ and $\mathbf Z_{-j}$ that contain all the observations except the ones in the $i$-th fold and denote them by $\mathbf Z_{j}^{(-i)}$ and $\mathbf Z_{-j}^{(-i)}$, respectively;
        \item for a given $\lambda_j^{\Theta}$, solve the minimization problem in \eqref{eq:gamma:U:Lasso} by using $\mathbf Z_{j}^{(-i)}$ and $\mathbf Z_{-j}^{(-i)}$:
            $$\wh\xi_{j}^{-i}(\lambda_j^{\Theta}) = \argmin_{\xi\in\mathbb{R}^{q-1}}\left\{\|\mathbf Z_j^{(-i)} - \mathbf Z_{-j}^{(-i)}\xi\|_2^2/n + 2\lambda_j^{\Theta}\|\xi\|_1\right\}.$$
            This gives $\wh\xi_{j}^{-i}(\lambda_j^{\Theta})$;
        \item compute the mean squared error $MSE^{(-i)}(\lambda_j^{\Theta})$ associated to the given $\lambda_j^{\Theta}$ as:
            \begin{displaymath}
               MSE^{(-i)}(\lambda_j^{\Theta}) := \|\mathbf Z_j^{(i)} - \mathbf Z_{-j}^{(i)}\wh\xi_{j}^{-i}(\lambda_j^{\Theta})\|_2^2/n.
            \end{displaymath}
    \end{enumerate}
    \item Compute the $10$-fold cross-validation estimate for the test mean squared error as
        \begin{equation}
          CV_{10}(\lambda_j^{\Theta}) = \frac{1}{10}\sum_{i=1}^{10} MSE^{(-i)}(\lambda_j^{\Theta}).
        \end{equation}
    \item Choose the $\lambda_j^{\Theta}$ that minimizes $CV_{10}(\lambda_j^{\Theta})$.
  \end{itemize}

\indent The cross-validation procedure to choose $\lambda_j^M$, $j=1,\ldots, p$ is the same as the one described in \textsc{Algorithm 1} with the following modification of steps 3-4 in the for loop:
  \begin{enumerate}
    \item[3.] for a given $\lambda_j^M$, solve the minimization problem in \eqref{eq:gamma:M:Lasso}:
      $$\wtl\gamma_{j}^{(-i)}(\lambda_j^M) = \argmin_{\gamma\in\mathbb{R}^{p-1}}\left\{\|((\wh \Theta^{1/2})^{(-i)}\wh M)_j - ((\wh \Theta^{1/2})^{(-i)}\wh M)_{-j}\gamma\|_2^2 + 2\lambda_j^M\|\gamma\|_1\right\}.$$
      This gives $\wtl\gamma_{j}^{(-i)}(\lambda_j^M)$;
    \item[4.] compute the mean squared error $MSE^{(-i)}(\lambda_j^M)$ associated to the given $\lambda_j^M$ as:
        \begin{displaymath}
          MSE^{(-i)}(\lambda_j^M) := \|((\wh \Theta^{1/2})^{(i)}\wh M)_j - ((\wh \Theta^{1/2})^{(i)}\wh M)_{-j}\wtl\gamma_{j}^{(-i)}(\lambda_j^M)\|_2^2.
        \end{displaymath}
  \end{enumerate}

\indent Finally, the cross-validation procedure to select the constant $c_n := C_0\sqrt{\log(q)/n}$ for the construction of $\wh M$ is given in the following algorithm.\\

\textsc{Algorithm 3.}
  \begin{itemize}
    \item For $i=1,\ldots,10$:
      \begin{enumerate}
        \item randomly select a set of observations in $\{(X_1,Z_1),\dots,(X_n,Z_n)\}$ of size $n_{tr} = \lceil n(1 - 1/log(n)) \rceil$. This is the training dataset and is denoted with a $tr_i$ index, the remaining observations will be the validation data denoted with a $v_i$ index;
        \item construct the submatrices of $\mathbf{X}$ and $\mathbf Z$ that contain only the observations in the validation fold and denote them by $\mathbf X^{(v_i)}$ and $\mathbf Z^{(v_i)}$, respectively;
        \item construct the submatrices of $\mathbf{X}$ and $\mathbf Z$ that contain only the observations in the training dataset and denote them by $\mathbf X^{(tr_i)}$ and $\mathbf Z^{(tr_i)}$, respectively;
        \item construct $\widetilde{M}^{(tr_i)} := (\mathbf Z^{(tr_i)})^T\mathbf X^{(tr_i)}/n_{tr}$;
        \item for a given $c_n$, compute the thresholding estimator by using \eqref{eq_M_est_thresholding}, $\mathbf X^{(tr_i)}$ and $\mathbf Z^{(tr_i)}$:
            $$\wh{M}_{jk}^{(i)}(c_n) := \widetilde{M}_{jk}^{(tr_i)} \1\left\{|\widetilde{M}_{jk}^{(tr_i)}| \geq c_n\right\}.$$
            This gives $\wh{M}_{jk}^{(i)}(c_n)$;
        \item compute the Frobenius norm $\|\cdot\|_F$ of the difference between $\wh{M}_{jk}^{(i)}(c_n)$ and $\widetilde M^{(v_i)} := (\mathbf Z^{(v_i)})^T\mathbf X^{(v_i)}/(n - n_{tr})$:
            \begin{displaymath}
               Loss^{(i)}(c_n) := \|\wh{M}_{jk}^{(i)}(c_n) - \widetilde M^{(v_i)}\|_F.
            \end{displaymath}
    \end{enumerate}
    \item Compute the mean of the losses as
        \begin{equation}
          Loss_{10}(c_n) = \frac{1}{10}\sum_{tr=1}^{10} Loss^{(i)}(c_n).
        \end{equation}
    \item Choose the $c_n$ in a grid that minimizes $Loss_{10}(c_n)$.
  \end{itemize} 

 \bibliography{BiB}
\end{document}